\documentclass[12pt]{article}
\usepackage{amsmath}
\usepackage{epsf}
\usepackage[scanall]{psfrag}
\usepackage{graphicx}

\makeatletter
\newcounter {newsection}[section]
\renewcommand \thenewsection {\@arabic\c@section}
\renewcommand \thesubsection {\@arabic\c@newsection.\arabic{subsection}}
\newcounter {newnewsection}[section]
\renewcommand \thenewnewsection {\@Alph\c@section}
\makeatother

\def\C{\mathbf{C}}
\def\F{\mathbf{F}}
\def\J{\mathbf{J}}
\def\boldg{\mathbf{g}}
\newcommand{\M}{\mathbf{M}}
\newcommand{\N}{\mathbf{N}}
\newcommand{\ran}{\rangle}
\newcommand{\lan}{\langle}
\newcommand{\flavor}{\mathrm{flavor}}
\newcommand{\diag}{\mathrm{diag}}

\def\U{\mathbf{U}}
\def\V{\mathbf{V}}
\def\Q{\mathbf{Q}}
\def\I{\mathbf{I}}
\def\X{\mathbf{X}}
\def\R{\mathbf{R}}

\def\fv{\F(v)}
\def\fop{\F(\mathrm{op})}

\def\lan{\langle}
\newcommand{\bfsigma}{\boldsymbol{\sigma}}

\newcommand{\bftheta}{\boldsymbol{\theta}}

\newcommand{\gev}{\mathrm{GeV}}
\newcommand{\tev}{\mathrm{TeV}}
\newcommand{\color}{\mathrm{color}}

\begin{document}

%\begin{flushright}
%\today\\
%DRAFT\\
%\end{flushright}
%\vskip-.5in

\begin{center}
{\Large Hidden Vector `Coordinates' in Particle Physics}\\[3mm]
\hskip1in\vbox{\hbox{\large Gerald L. Fitzpatrick}
\hbox{\em PRI Research and Development Corp.}
\hbox{\em 12517 131 Ct. NE}
\hbox{\em Kirkland, WA 98034}
\hbox{glfitzpatrick@yahoo.com}}
\end{center}

\begin{abstract}
\noindent Heretofore unrecognized (i.e., ``hidden'') Lorentz-invariant vector \emph{observables} in the fermion sector (i.e., flavor-defining fermion ``coordinates'')  are shown to indirectly explain most, if not all, of the so-called ``accidental'' (internal) symmetries associated with fundamental fermions (quarks and leptons), by explaining quark and lepton \emph{flavors}, \emph{flavor doublets} and \emph{families}. Moreover, these new fermion coordinates lead to \emph{quantitative} constraints on neutrino mixtures that are found to be in good agreement with current experimental observations. The new fermion coordinates arise in the context of a \emph{quantized}, internal 2-space possessing a non-Euclidean metric. ``Quantization'' here means that all \emph{scalars}, \emph{vectors} and \emph{matrices} relevant to a description of fundamental fermions in the 2-space, are both \emph{discrete} and \emph{finite} in number.  In a certain limited sense, the new 2-space description of fundamental fermions is more fundamental, and more general, than that provided by the standard model of particle physics. In particular, it points to  a ``layer'' of new physics---located \emph{somewhere} between the unification scale, and the region of applicability of the standard model---that is responsible for the 2-space, its quantization, and its associated ``selection rules.''
\end{abstract}

\section{Introduction}

Most physicists would probably agree that the standard model of particle physics [1--3], based on the gauge group  $SU(3)_\color\times SU(2)_L\times U(1)_Y$, is only a low-energy or ``macroscopic'' approximation to a deeper, more fundamental, \emph{unified} theory, possibly involving either Planck level physics at an energy scale of $10^{19}\gev$ \cite{4,5}, or a much lower $\tev$ energy scale should the anticipated \emph{extra} dimensions turn out to be large \cite{6}. Consider for example, the case of the standard-model Lagrangian.

Given the number of flavors of quarks and leptons, and an appropriate (renormalizable) strong-electroweak Lagrangian, the so-called ``accidental symmetries'' of the Lagrangian \cite{7} are known to ``explain'' both the existence, and separate (exact or approximate) conservation, of various (global) ``charges'' associated with quarks and leptons (e.g., \emph{lepton number, baryon number, strangeness, charm, truth, beauty, electron-, muon-} and \emph{tau-numbers}). However, \emph{there is nothing in such Lagrangians, or their associated accidental symmetries, that would explain quarks and leptons, or tell us how many} flavors \emph{of quarks and leptons to include in the Lagrangian}. Rather, it seems that the Lagrangian is \emph{not} an explanation for these things, but instead is a \emph{result} of physics at some deeper, more fundamental, level, possibly involving \emph{unification}. However, if unification really does occur, we are immediately faced with the important unanswered question of exactly how the standard model of particle physics, and its associated Lagrangian, arises or \emph{emerges} from this deeper, more fundamental, ``substratum.''

In this paper  these questions are addressed in a new way by identifying what may turn out to be an intermediate ``layer'' of new physics located \emph{somewhere} between the unification scale (wherever it happens to be), and the region of applicability of the standard model of particle physics. The strategy pursued 
here is similar to that of S.\ Goudsmit and G.\ Ulenbeck \cite{8}, who
first
realized that a description of the observed \emph{fine structure} in atomic
spectra was \emph{not} possible unless electrons possessed, in addition to their
mass and electric charge, a previously hidden \emph{spin} ``coordinate.'' In
particular, 
the principal purpose of this paper is to argue that the proposed intermediate ``layer'' of new physics is described by heretofore unrecognized (i.e., ``hidden'')  2-vector observables  in the fermion sector (i.e., flavor-defining fermion ``coordinates'' $\U$, $\V$ and $\Q$), which serve to explain, among many
other
things, the proliferation of matter fields (flavors) in comparison to the relative paucity of force-mediating particles. 
And, because we will also argue here 
that the new flavor-defining fermion coordinates (i.e., the 2-vectors $\Q$, $\U$, $\V$ where $\Q=\U+\V$) are \emph{Lorentz-invariant}, we will be free to perform a variety of 
physically meaningful \emph{algebraic}, \emph{geometric}, and \emph{topological} manipulations on any one (or more) of these coordinates---\emph{without undue concern for how these manipulations ``appear'' to observers located in arbitrary} (\emph{inertial}) \emph{coordinate systems}.

From the foregoing, it is clear that 
once new coordinates of this kind
 are introduced in the fermion sector, an entirely new description of fundamental fermions (quarks and leptons) 
must result. Moreover, the 
proposed description appears to be both more fundamental, and more general (in certain limited respects), than the standard model description of fundamental fermions. For example, by employing these new fermion coordinates, the question  of
why fundamental fermion families are \emph{replicated}, and the question of why there are just \emph{three}  families, can
both be answered, whereas this is \emph{not} possible in the standard model description [1--3]. Thus the new flavor-defining fermion coordinates serve both to \emph{complement}, and to (modestly) \emph{extend}, the standard model of particle physics.

This work is based, in part, on previous works by the author wherein different aspects of this approach to flavors, families, and the problem of family replication, were discussed [9--13]. These works provide a different emphasis, and in some cases, a more detailed description of many features of this general idea that are only touched upon in the present work, which  essentially  \emph{reviews,
clarifies}, and \emph{consolidates} these earlier works.

\section{The Electric Charge Coordinates Q}

It is well known that in quantum mechanics the wave function or state vector $\Psi_n$ describing an individual (``isolated'') particle, provides all of the information that is available, in principle, about that particle. Here, the subscript on $\Psi_n$ refers to various kinds of \emph{simultaneously observable} quantum numbers associated with the particle. Hence, it is possible to effectively define such a quantum state by specifying the specific values of these simultaneous observables \cite{14}. In a similar way,  there are known to exist certain simultaneously observable flavor-defining quantum numbers (e.g., strangeness and baryon number) that serve to define individual flavor or antiflavor eigenstates of quarks and leptons [15, 16]. It is proposed here that these flavor-defining quantum numbers are to be identified with certain global charge-conjugation-reversing ($\C$-reversing) scalars, which are \emph{derived}, in turn, from the  flavor-defining fermion coordinates, $\Q$, $\U$ and $\V$.

If ever there was a signal for new coordinates in particle physics, surely it comes from the known fundamental-fermion sector. Unlike the known ``force-mediating'' fundamental-boson sector (i.e., the \emph{photon, gluons, \emph{the} graviton, \emph{and} weak intermediate vector bosons}), matter-particles or fundamental fermions (quarks and leptons), exhibit a relative plethora of \emph{flavors} falling into  three families. Counting only color-flavor states, there are 48 known fundamental fermions and antifermions \cite{17}, compared with only 13 force-mediating bosons \cite{18}. What could be more natural than to assume that this ``embarrassment of riches'' in the fermion sector, is due to new flavor-defining fermion coordinates, which are \emph{not} carried by the force-mediating bosons \cite{18, 19}?

In a recent paper \cite{10}, it was shown that the fermion-number operator $\fop$ itself may be the key to a preliminary understanding of how new flavor-defining fermion coordinates such as the electric-charge coordinate $\Q$ arise. In particular, let us review the salient points in \cite{10} to see how these coordinates arise from an \emph{analytic continuation} of the fermion-number operator $\fop$.

\setcounter{newsection}{2}

\subsection{Analytic continuation of F(op) and a new internal non-Euclid-
ean 2-space}

We have found that an analytic-continuation of a Hermitian matrix $\fop$ representing
the conventional fermion-number
\emph{operator}, from an \emph{external} spacetime and Hilbert-space setting,
to a new \emph{internal} (real) non-Euclidean space---$\fop$ is continued to a real, generally  non-diagonal matrix 
$\fv$ involving a single \emph{real} parameter $v$---``automatically'' leads to a new description of fundamental fermions (quarks and leptons) in
which families are \emph{replicated} and there are just \emph{three} families. The fact that this happens,
suggests that there is some deep connection between the result of the aforementioned 
analytic-continuation, namely $\fv$, and some more fundamental underlying physics (e.g., Planck-level physics) where flavor degrees-of-freedom, and family-replication, presumably originate.

\subsubsection{The  fermion-number operator F(op)}

Consider the situation, presumably at some high energy, where we are dealing with ``free'' (isolated) leptons or ``analytically free'' quarks. Suppose we want to describe the scalar fermion-number carried by these particles. And, suppose further, that the energies involved are not so high that quantum field-theory  (QFT) breaks down. Under these conditions the fermion-number can be represented by  a $U(1)$-type scalar ``charge'' \cite{20, 21}, namely, a charge associated with the (continuous)
group of  unitary matrices $U$ of order 1 known as $U(1)$. 

The
fermion-number operator, which can be represented by a Hermitian matrix $\fop$, is said to generate
these so-called ``gauge'' (or phase) transformations, which in turn act on
fermion and antifermion quantum states in Hilbert space. That is,
given that $\alpha$ is a \emph{real} phase  one has
\begin{equation}
U=e^{i\alpha\fop},
\end{equation}
and for infinitesimal transformations [i.e., $e^{i\,\delta\alpha\,\fop}=
1+i\,\delta\alpha\,\fop$] acting on single-particle (free or ``asymptotically free'')
fermion and antifermion states
$|p\rangle$ and $|\overline p\rangle$, respectively, one easily
establishes that (the fermion-number ``charges'' are  $f_m=-f_a=1$ for matter and $f_a$ for antimatter)
\begin{equation}
\begin{array}{rcl}
U|p\rangle & =  e^{i\,\delta\alpha\,f_m}|p\rangle \\
U|\overline p\rangle & = e^{i\,\delta\alpha\,f_a}|\overline p\rangle, 
\end{array}
\end{equation}
since, by definition, $\fop$ obeys the eigenvalue equations
\begin{equation}
\begin{array}{rcl}
\fop|p\rangle & = f_m|p\rangle \\
\fop|\overline p\rangle & =f_a|\overline p\rangle.
\end{array}
\end{equation}

Finally, the assumption that the Hamiltonian  $H$ is invariant under $U$, namely
\begin{equation}
H=UHU^\dagger,
\end{equation}
ensures that $H$ and $\fop$ commute
\begin{equation}
[\fop, H]=0,
\end{equation}
as can be verified by differentiating $UHU^\dagger$ with respect to
$\alpha$. Hence, the total fermion-number (the number of \emph{fermions} minus the
number of \emph{antifermions}) is a constant of the motion.

\subsubsection{Matrix representation of the fermion-number operator F(op)}

Because the matrix $\fop$ involves just \emph{two} kinds of
quantum states (3),
namely $|p\rangle$ and $|\overline p\rangle$,  it can be expressed as a $2\times 2$ diagonal Hermitian matrix (6), where one of the 
adjustable parameters $(\theta)$ is a \emph{fixed} constant (up to $2\pi$)
and the other $(\phi)$ is freely \emph{adjustable}. In particular, 
\begin{equation}
\fop = \left.\left( \begin{array}{ll}
\cos\theta & \sin\theta e^{-i\phi} \\
\sin\theta e^{+i\phi} & -\cos\theta
\end{array}\right)\right|_{\cos\theta=1}=\sigma_z,
\end{equation}
where 
\begin{equation}
\sigma_z=\left(\begin{array}{cc}
f_m & 0 \\
0 & f_a\end{array}\right)
\end{equation}
is one of the familiar Pauli matrices.

This form for $\fop$ is consistent with (3) 
where the normalization and orthogonality conditions, namely $\langle p|p\rangle = \langle\overline p|\overline p\rangle=1$ and $\langle p|\overline p\rangle = \langle\overline p|p\rangle =0$, respectively, directly yield
\begin{equation}
\fop = \left( \begin{array}{ccc}
\langle p|\fop|p\rangle & , & \langle p|\fop|\overline p\rangle
\\
\langle\overline p|\fop|p\rangle & , & \langle\overline
p|\fop|\overline p\rangle
\end{array}\right)=\sigma_z.
\end{equation}
Note that owing to (7) and (8), $\cos\theta<1$ in (6) is excluded.
Here it should also be noted that $tr\fop=f_m+f_a=0$, $det\fop=f_m\cdot f_a=-1$, and $\mathbf{F}^2\mathrm{(op)}=\I_2$ is the $2\times 2$ identity matrix.

\subsubsection{The Continuation From F(op) to a `generalized fermion-number' F($v$)}

Now perform an analytic continuation on $\fop$, namely $\fop\to\fv$, which maintains $\fv$ \emph{real} and $\cos\theta\ge 1$. This can only be accomplished by continuing $\theta$ from a  \emph{real} to an \emph{imaginary} number, and by maintaining $e^{-i\phi}$  \emph{imaginary}. In particular, to maintain $\fv$ real, we must make the replacements $\theta\to iv$ and $e^{-i\phi}\to \mp
i$, where $v$ is a \emph{real} number. Then
\begin{eqnarray}
\fv & = & \left.\left(\begin{array}{ll}
\cos\theta & \sin\theta e^{-i\phi} \\
\sin\theta e^{+i\phi} & -\cos\theta\end{array}\right)\right|_{\theta=iv, e^{-i\phi}=\mp
i}\\
&\hbox{or}&\nonumber \\
\fv & = & \left(\begin{array}{rl}
\cosh v & \pm \sinh v \\
\mp\sinh v & -\cosh v \end{array}\right), 
\end{eqnarray}
where, just as for $\fop$, the eigenvalues of $\fv$ are $f_m$ and $f_a$, and so we have $tr\fv=f_m+f_a=0$, $det\fv=f_m\cdot f_a=-1$, and $\F^2(v)=\I_2$.

In Ref.\ 9, pp.\ 50 and 54 it is argued that only the upper signs in (10) have physical significance and $v\ge 0$.
And, just what is the physical significance of $\fv$?

Because the continuation ``connects'' $\fop$ and $\fv$, it is natural to assume that \emph{both} $\fop$ and $\fv$  describe, or represent, aspects of the fermion number (i.e., the matter-antimatter ``degree-of-freedom'' or matter-antimatter ``dichotomy''). However, unlike $\fop$ or $f$, $\fv$ is a kind of ``generalized fermion-number,'' which will be shown to describe additional degrees-of-freedom such as the ``up''-``down'' and quark-lepton degrees-of-freedom.
Moreover, it is abundantly clear from (10) that the generally non-Hermitian (when $v\ne 0$) matrix $\fv$---unlike the Hermitian matrix $\fop$ in (8), which acts on Hilbert space---does \emph{not} act on a Hilbert space in an \emph{external} spacetime setting
\cite{22}.

\subsubsection{A new internal  non-Euclidean 2-space}

Because of the matter-antimatter \emph{dichotomy} exhibited by fundamental fermions, the fermion-number operator $\fop$ is necessarily a 2 by 2 matrix, and so the continuation of $\fop$, namely $\fv$, is also a 2 by 2 matrix. Therefore, \emph{the matrix} $\fv$ \emph{must} ``\emph{act}'' \emph{on some new internal} 2-\emph{dimensional space}. What is the nature of this space? In particular, assuming that it is a metric space, what is the metric?

When the matrix $\fv$ acts on a real column-vector $\{a, b\}$, it leaves the
quadratic form
$a^2-b^2$, invariant.  Therefore, the 2-space metric is non-Euclidean or
``Lorentzian'', and can be represented by the matrix \cite{23}
\begin{equation}
\boldg = \left(
\begin{array}{rr}
1 & 0 \\
0 & -1 \end{array}\right).
\end{equation}
Given this metric, the \emph{scalar product} of two real vectors (i.e., the ``projection'' of one vector upon another) assumes the
form
\begin{equation}
(a,b)\{e,f\}=ae-bf.
\end{equation}
Similarly,  the \emph{square} of a real vector (i.e., the ``projection'' of the vector upon itself) is given by
\begin{equation}
(a,b)\{a,b\}=a^2-b^2.
\end{equation}
In (12) and (13) above, $(\;,\;)$ is a \emph{row} vector while $\{\;,\;\}$ is a (conformable)
\emph
{column} vector.

\subsection{Quantization of the imaginary angle $\bftheta$ and the electric-charge coordinates Q}

It is very important to understand that the imaginary angle $\theta$ in (9) must be \emph{quantized}. That is, when we do the continuation to an imaginary angle, the only allowed values seem to be
\begin{equation}
\theta=iv=i\ln M_c,
\end{equation}
or
\begin{equation}
e^{-i\theta}=M_c,
\end{equation}
where $M_c=1$ or 3 only (see Refs.\ 24, 25, and Ref 9, p.\ 79).

This quantization of $\theta$ and/or $v$ comes about in the following way. Because the eigenvalues of $\fv$ are $\C$-reversing quantized \emph{charges}, the components of both the diagonal and off-diagonal forms of the matrix $\fv$ are, necessarily, \emph{quantized} charges. Since the components of $\fv$ involve hyperbolic functions of $v$, it follows that $e^v$, $e^{-v}$ and $v$ are, necessarily \emph{quantized}. In particular, as we  show elsewhere $fe^{-v}=f/M_c$ is a quantized charge identified with the \emph{baryon} or \emph{lepton} number (Ref.\ 9, pp.\ 52, 53), depending on the value of the quantum number $M_c$. And, from the spin-statistics theorem, we know that an \emph{odd} number of fundamental fermions must bind together (``strongly'') to form a baryon or lepton. Hence, the quantum number $M_c$ must be an odd integer $M_c=1, 3, 5, \ldots$. (see Refs.\ 24, 25, and Ref.\ 9, p.\ 79).

\begin{center}
\begin{psfrags}
\includegraphics[width=4in]{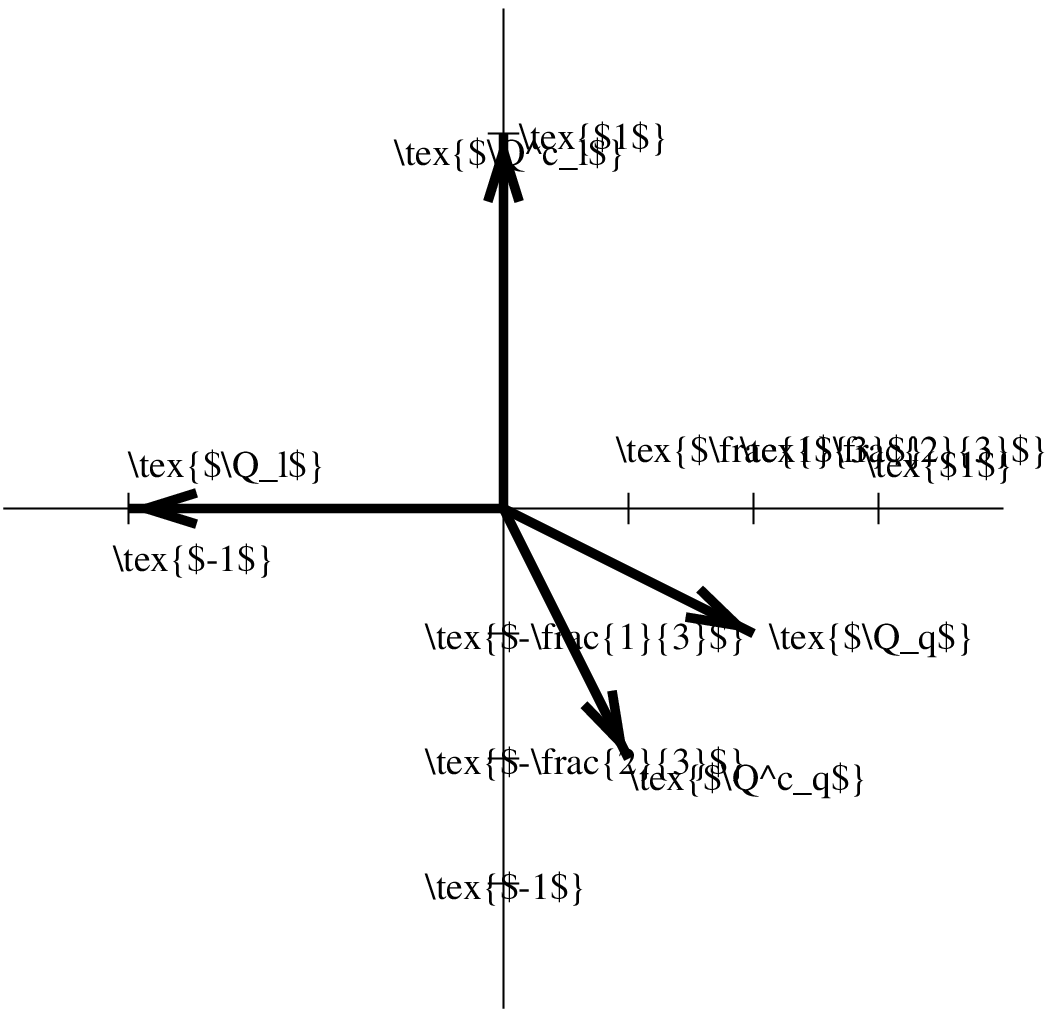}
\end{psfrags}
\end{center}
\noindent{\bf Figure 1}.
{The \emph{four} ``electric
charge'' 2-vector fermion and antifermion ``coordinates'' $\Q_q, \Q^c_q,
\Q_l, \Q^c_l$.}
\bigskip

These identifications are remarkable for at least two reasons. First, we discover that the eigenvectors of $\fv$, and related functions evaluated at $M_c=1$ and $M_c=3$, yield the quantized \emph{electric} charges of both leptons and quarks, and their associated quantized \emph{lepton} or \emph{baryon}  numbers, respectively \cite{26}. In particular, choose the upper signs in (10),
\begin{equation}
\F(v) = \left( \begin{array}{rr}
\cosh v & \sinh v\\
-\sinh v & -\cosh v
\end{array}\right).
\end{equation}
Then  identify (see Ref.\ 9, pp.\ 52--55)
 the quark and lepton electric charges with the 
``up''-``down'' components of the eigenvectors of the matrix $\fv$. That is, the quark charges are given by $(M_c=3)$
\begin{eqnarray}
q_1(f) & = & \frac{(M^2_c-1)}{2M_c(M_c-f)} = +\frac{2}{3}\hbox{ for }f=+1\hbox{ and }+\frac{1}{3}\hbox{ for }f=-1,  \\
q_2(f) & = & q_1(f)-1, 
\end{eqnarray}
where the baryon number for quarks is the non-Euclidean quadratic form $\Q^2_q= B=q^2_1(f)-q^2_2(f)=\pm\frac{1}{3}$ for $f=\pm 1$, where the quark electric-charge ``coordinate'' (see Fig.\ 1) $\Q_q=\Q_q(f,M_c)=\{q_1(f), q_2(f)\}$ is an eigenvector of $\fv$.
Similarly, the lepton electric charges are given by $(M_c=1$)
\begin{eqnarray}
q'_1(f) & = & \frac{-(M_c^2-1)}{2M_c(M_c-f)} = -1\hbox{ for }f=+1\hbox{ and }0\hbox{ for }f=-1,  \\
q'_2(f) & = & q'_1(f)+1, 
\end{eqnarray}
where the lepton number for leptons is the non-Euclidean quadratic form $\Q^2_l=L=[q'_1(f)]^2-[q'_2(f)]^2=\pm 1$ for $f=\pm 1$, where the lepton electric-charge ``coordinate'' (see Fig.\ 1) $\Q_l=\Q_l(f,M_c)=\{q'_1(f), q'_2(f)\}$ is an eigenvector of $\fv$.

Second, it is abundantly clear that the quantum number $M_c$ can be identified with the strong-color ``multiplicity.'' In particular, \emph{leptons} being strong-color \emph{singlets} $(M_c=1)$ do not bind strongly to form a composite particle, whereas \emph{quarks} being strong-color \emph{triplets} $(M_c=3)$, do bind strongly to form a composite baryon. That is, $M_c$ is both a measure of the strong-color multiplicity and the \emph{number} of quarks or leptons that bind (strongly) to form composite particles.

It is remarkable that the 2-space description of fundamental fermions has
``automatically' yielded, not only the \emph{electric, baric} and \emph{leptic}
charges of quarks and leptons, but also an unanticipated connection with
\emph{quantum
 chromodynamics} (QCD) and $SU(3)_\color$. In particular, the 2-space and
the associated quantized parameter $v=\ln M_c$, implies that something very
similar to
 $SU(3)_\color$ must be a symmetry of nature. \emph{We take this
circumstance to
 be a strong indication that the 2-space and/or whatever underlying physics
is responsible for it, is more primitive} (\emph{i.e., more fundamental})
\emph{than
 the standard model of particle physics.}

\subsubsection{Representing charge conjugation C in the 2-space}

Given that there are numerous $\C$-reversing scalars such as $q_1, q_2, q'_1, q'_2$, $B$ and $L$, in the 2-space, there must exist a $2\times 2$ matrix, call it $\X$, that serves to transform these \emph{scalars}, various 2-\emph{vector} fermion coordinates (e.g., $\Q$, $\U$ and $\V$), and various $2\times 2$ \emph{matrices} such as $\fv$, to their corresponding $\C$-reversed (2-space) counterparts. For example, a matrix $\X$ should exist such that
\begin{equation}
\X\Q=\Q^c
\end{equation}
and
\begin{equation}
\X\Q^c=\Q.
\end{equation}
From (21) and (22) it follows that $\X$ must equal its multiplicative inverse $(\X=\X^{-1})$, and thus
\begin{equation}
\X^2=\I_2,
\end{equation}
where $\I_2$ is the $2\times 2$ identity matrix.

Write $\X$ in the general form ($\X$ is real)
\begin{equation}
\X = \left(\begin{array}{cc}
a & b \\
c & d \end{array}\right),
\end{equation}
and consider the situation for leptons (see Eqs.\ 19 and 20), where one of the two charges ($q'_2$) associated with $\Q_l=\{q'_1(1)=-1=q'_1$, $q'_2(1)=0=q'_2\}$ and $\Q^c_l=\{q'_1(-1)=0=q'_2$, $q'_2(-1)=+1=-q'_1\}$ is \emph{zero}, and the other charge $(q'_1)$ is \emph{nonzero}.

In this particular case, we have
\begin{equation}
\left(\begin{array}{cc}
a & b \\
c & d \end{array}\right)\left(\begin{array}{c}
q'_1 \\
0 \end{array}\right) = \left(\begin{array}{c}
0 \\
-q'_1 \end{array}\right),
\end{equation}
which means that
\begin{equation}
aq'_1=0
\end{equation}
and
\begin{equation}
cq'_1=-q'_1.
\end{equation}

And therefore, since $q'_1\ne 0$, it must be true from (26) and (27) that $a=0$ and $c=-1$.

Since 
$\X\Q^c=\Q$ it must also be true that
\begin{equation}
\left(\begin{array}{rr}
0 & b \\
-1 & d \end{array}\right)\left(\begin{array}{c}
0 \\
-q'_1 \end{array}\right) = \left(\begin{array}{c}
q'_1 \\
0 \end{array}\right),
\end{equation}
which means that
\begin{equation}
-bq'_1 = q'_1 
\end{equation}
and
\begin{equation}
-dq'_1 = 0.
\end{equation}
Finally, since $q'_1\ne 0$ it must be true from (29) and (30) that $b=-1$ and $d=0$.

Collecting the foregoing matrix elements, we have
\begin{equation}
\X = \left(\begin{array}{rr}
0 & -1 \\
-1 & 0 \end{array}\right),
\end{equation}
or
\begin{equation}
\X=-\bfsigma_\X,
\end{equation}
where $\bfsigma_\X$  is one of the familiar Pauli matrices. In general, the matrix $\X=-\bfsigma_\X$ should apply (in the 2-space) to 2-\emph{scalars}, 2-\emph{vectors},
$2\times 2$ \emph{matrices}, and to both \emph{quarks} and \emph{leptons}.

\subsubsection{Transformations of charge-like scalars associated with Q, U and V under X}

As demonstrated in Section 2.2.1, the charge-like scalar \emph{components} of 2-vector coordinates such as $\Q_q$ or $\Q_l$ change signs under $\X$. However, owing to the 2-space metric given by (11), scalar products of 2-vectors whose components transform like charges, will also be charge like.

For example,  we  have the \emph{square} of the 2-vectors $\Q$ and $\Q^c$, namely,
\begin{equation}
\Q^2=\Q\bullet \Q=q^2_1-q^2_2
\end{equation}
and 
\begin{equation}
(\Q^c)^2 = \Q^c\bullet \Q^c=q^2_2-q^2_1.
\end{equation}
Therefore,
\begin{equation}
\Q^2= -(\Q^c)^2,
\end{equation}
which means that $\Q^2$ and $(\Q^c)^2$ each transform like $\C$-reversing 2-scalar \emph{charges}. 

As described previously, these particular charges can be identified with the \emph{baryon}- or \emph{lepton}-\emph{number} carried by quarks or leptons, respectively (see Ref.\ 9, p.\ 72).
We
will see in a later section that when charge vectors such as $\Q$ (or $\Q^c$) are \emph{resolved} in the 
2-space into pairs of linearly independent vectors (e.g., $\Q=\U+\V$), not only are the \emph{components} of $\Q$, $\U$ and $\V$,
$\C$-reversing \emph{charges}, but also by squaring $\Q=\U+\V$, namely,
\begin{equation}
\Q^2=\U^2+2\U\bullet \V+\V^2,
\end{equation}
$\U^2$, $2\U\bullet \V$ and $\V^2$ are, like $\Q^2$, $\C$-reversing charges. The foregoing
collection of 2-scalar charges will be used to define and describe \emph{flavor eigenstates}, \emph{flavor doublets}, and eventually \emph{families} of fundamental fermions \cite{27, 28}.

\subsubsection{Transformation of the metric and other matrices under $\X$}

Any $2\times 2$ matrix $\M$, appropriate to the 2-space description of fundamental fermions, should transform under $\X=-\bfsigma_\X$ to its $\C$-reversed counterpart $\M^c$ according to the \emph{similarity transformation}
\begin{equation}
\X\;\M\;\X^{-1}=\M^c,
\end{equation}
or because $\X=\X^{-1}=-\bfsigma_\X$, equivalently as
\begin{equation}
(-\bfsigma_\X)\M(-\bfsigma_\X)=\M^c.
\end{equation}
For example, the metric $\boldg$ (see Eq.\ 11) is found to be $\C$-\emph{reversing} since
\begin{equation}
(-\bfsigma_\X)\boldg(-\bfsigma_\X)=-\boldg.
\end{equation}
Similarly, the matrix $\fv$ in (16) is $\C$-reversing since $(-\bfsigma_\X)$  $\fv(-\bfsigma_\X)=-\fv$.

A matrix that is $\C$-\emph{invariant} (e.g., the matrix $\X$) would, necessarily, have the form
\begin{equation}
\N=\left(\begin{array}{cc}
a& b \\
b & a \end{array}\right),
\end{equation}
where it is clear that
\begin{equation}
(-\bfsigma_\X)\N(-\bfsigma_\X)=\N.
\end{equation}

\section{What Dynamics Governs the Evolution of Flavor-Defining Fermion Coordinates Q, U and V?}

As a first approximation, it is safe to say that the \emph{dynamics} of a system consisting of an ``isolated'' or individual elementary particle, is governed by the \emph{evolution} of the system wave function or state vector. The state vector in turn evolves according to quantum mechanical equations of motion that can be said to arise from an appropriate strong-electroweak Lagrangian (when the standard model applies) or a more fundamental Lagrangian in case deeper physics applies \cite{29}. The question that naturally arises in the present situation is how such a dynamical description is altered or constrained, if at all, by the new flavor-defining fermion coordinates $\Q$, $\U$ and $\V$? Even a partial answer to this question may suggest ways that the new 2-space description of fundamental fermions can be experimentally tested. Let us begin the discussion by describing how quantum mechanical state vectors can be partially represented using the new 2-vector fermion coordinates $\Q$, $\U$ and $\V$.

\setcounter{newsection}{3}
\subsection{Representing flavor eigenstates and flavor doublets in the 2-space}

The continuation from the 2-D
 Hilbert space to the 2-D non-Euclidean ``charge'' space (see Sec.\ 2.1.3) turns out to
mean that individual flavors of fundamental fermions 
can be partially  represented in an unconventional way by geometric objects (in the non-Euclidean 2-space) 
which \emph{differ} from a quantum state, but \emph{from which the quantum states 
can be inferred or effectively constructed}.  In particular, in the non-Euclidean 2-space, an object 
we call a ``vector triad" (i.e., the ``triad'' of flavor-defining fermion coordinates $\Q$, $\U$ and $\V$) represents ``up"-`` down" type flavor \emph{doublets} of
fundamental fermions and antifermions---the ``up"-``down" type flavor dichotomy.  That is, 
the components of the vectors associated with a given vector-triad 
are \emph{observable} Lorentz-invariant (Lorentz 4-scalar) ``charges," which can be 
used to  define the \emph{two} flavor-eigenstates in a flavor doublet \cite{15, 27, 28, 16, 30}.

\subsubsection{Flavor eigenstates}

As demonstrated in detail in 
Ref.\ 9, pp.\ 16--18, given the charge-like ($\C$-reversing) \emph{observables} associated
with the description involving 
$\fv$, namely, the real $\C$-reversing scalar-components of various matrices
and
 vectors defined on 
the internal non-Euclidean 2-space, it is possible to write down flavor
eigenstates \cite{16, 30}.

In principle, what one does is to identify the mutually-commuting $\C$-reversing 
``charges" (call them $C_i$) or charge-like quantum numbers associated 
with a particular flavor, and then write the corresponding simultaneous 
flavor-eigenstate as
\begin{equation}
|C_1, C_2, C_3, \ldots, C_n\rangle.
\end{equation}
Here $C_1, C_2, C_3, \ldots, C_n$, are the ``good" charge-like flavor-defining quantum 
numbers (charges) associated with a particular flavor.  It happens that 
these \emph{observable}  real-numbers can be identified with quantum 
numbers such as \emph{electric charge, strangeness, charm}, the 
third-component of (global) \emph{isospin, truth and beauty} (see Ref.\ 9, p.\ 72).
To discover what charges describe a particular flavor we
must first identify the vector-triad associated with that flavor.

Now, each \emph{vector-triad} represents a flavor \emph{doublet}, not just an individual
flavor.  That is,
vector-triads provide information on \emph{two} quantum states (two simultaneous
flavor-eigenstates)
associated with flavor doublets.  Therefore, vector-triads are associated with both
individual flavors \emph{and}
flavor doublets.  Here we simply summarize
how it is the that non-Euclidean vector-triads can be associated with  both
\emph{individual} flavors and ``up"-``down" type
\emph{flavor-doublets}.

\subsubsection{Flavor doublets}

Consider the eigenvector (call it $\Q$) of $\fv$ for fundamental fermions (see Footnote 26, and Sec.\ 2.2).
Since the space on which $\fv$
``acts" is two-dimensional, the \emph{observable} vector $\Q$ can
be ``resolved" into two (no more or less)
\emph{observable},  linearly-independent vectors, call them $\U$ and $\V$,
as
$\Q = \U + \V$ \cite{31, 32}.  Now, because these
three vectors ($\Q$, $\U$, and $\V$) are \emph{simultaneous}
observables, it makes sense to speak of this ``triad"
of vectors as being a well defined geometric object, namely, a ``vector triad."

Recognizing that the components of $\Q$, $\U$ and $\V$ are $\C$-reversing
charge-like \emph{observables} we can
write these \emph{observable} ``charge" vectors as
\begin{eqnarray}
\Q & = & \{q_1, q_2\} \\
\U & = & \{u_1, u_2\} \\
\V & = & \{v_1, v_2\},
\end{eqnarray}
where $q_1$, $q_2$, $u_1$, $u_2$, $v_1$ and $v_2$ are the various
\emph{observable} ``charges" (e.g., $q_1$ and $q_2$ were found to be \emph{electric} charges). 
Given $\Q = \U + \V$, the non-Euclidean metric (11), and Eqs.\ (43) through
(45), we
 find the 
associated \emph{observable} quadratic ``charges"   
\begin{eqnarray}
\Q^2 & = & \U^2 + 2 \U\bullet\V + \V^2 \\
2\U\bullet\V & = & 2(u_1v_1-u_2v_2) \\
\U^2 & = & u^2_1 - u^2_2 \\
\V^2 & = & v^2_1-v^2_2.
\end{eqnarray}
Finally, using the foregoing flavor-defining charges, we can express the \emph{two} quantum
states (simultaneous 
flavor-eigenstates) associated with a \emph{single} vector-triad in the
form of ``ket" vectors as follows (see Ref.\ 14, and Ref.\ 9, pp.\ 16--18)
\begin{equation}
\begin{array}{l}
|q_1, u_1, v_1, \Q^2, \U^2, 2\U\bullet \V, \V^2\rangle, \\
|q_2, u_2, v_2, \Q^2, \U^2, 2\U\bullet \V, \V^2\rangle.
\end{array}
\end{equation}
Here, the state $|q_1, u_1, v_1, \Q^2, \U^2, 2\U\bullet \V, \V^2\rangle$
 represents the ``up"-type flavor-eigenstate, and
$|q_2, u_2, v_2, \Q^2, \U^2, 2\U\bullet\V, \V^2\rangle$
 represents the corresponding ``down"- type flavor-eigenstate
in a flavor doublet of fundamental fermions \cite{16, 30}.

In this section we have demonstrated, among other things, that there are essentially two ways the new flavor-defining fermion coordinates $(\Q, \U, \V)$ can be \emph{combined}. In particular, when $(\Q, \U, \V)$ apply to the \emph{same} fundamental fermion, these vectors can either be \emph{added}, as in $\Q=\U+\V$, or they can be \emph{multiplied}, as in the non-Euclidean scalar products (i.e., ``projections'') expressed by Equations 46--49. However, as described in the next section, \emph{the linear superposition principle of quantum mechanics severely limits the ways} $(\Q, \U, \V)$ \emph{can be combined when \emph{two} or \emph{more} fundamental fermion flavors are involved.}

\subsection{When the new internal coordinates \emph{cannot} be superimposed}

According to the linear superposition principle of quantum mechanics (QM), wave functions or state vectors for two or more particles, and certain of their associated simultaneously observable charge-like quantum numbers or ``charges,'' can be \emph{superimposed}. And, because the overarching linear superposition principle of QM must be retained, it turns out that \emph{like} flavor-defining fermion coordinates associated with (``carried by'') \emph{two} or more fundamental fermions \emph{cannot} be superimposed or \emph{added} ($\Q$ with $\Q$, $\U$ with $\U$ and $\V$ with $\V$). For, if they could be superimposed, the linear superposition principle of QM could \emph{not} be retained, nor could flavors be defined in a self-consistent way using the fermion coordinates $(\Q, \U, \V)$.

In particular, one obvious reason that like coordinates ($\Q$, $\U$ or $\V$) for two or more fundamental fermions cannot be superimposed, is that such superpositions inevitably lead to nonsensical charge-like quantum numbers (``charges''), which could \emph{not} be ``carried'' by a sensible state vector, or used to define the corresponding wave function or state vector for the collection of particles in question. Let us demonstrate the validity of this assertion by giving a few simple examples, which serve to demonstrate some of the rules governing the superposition (or lack thereof) of the 2-vector coordinates $\Q$, $\U$ and $\V$.

\subsubsection{Rules for combining Q, U and V}

Consider the case of three quarks bound together by strong forces to form a baryon with baryon number $B=1$. Since the baryon number of each quark is given by $\Q^2=1/3$, the baryon number for three bound quarks is simply $3\Q^2=B=1$. Now this result makes it clear that we cannot simply add the three vectors $\Q$ together first, and then ``square'' them to get the baryon number for such a composite particle. In particular, it is clear that the baryon number of a composite consisting of three quarks is \emph{not} given by $(3\Q)^2=B=3$.

Now it might be supposed that this algorithmic failure can be attributed to the absence of an appropriate ``normalization'' factor when adding such vectors. For example, suppose that we were to make the rule that when $N$ of these $\Q$-vectors are added, we must divide the vector sum by $\sqrt{N}$. Then for the case of three quarks we would have a baryon number given by $(3\Q/\sqrt{3})^2=B=1$, which is certainly correct. However, the maximum, and minimum, electric charges of such a composite would then be given, respectively, by the scalar products $(3\Q/\sqrt{3})\{1,0\}=+2\sqrt{3}/3$ and $(3\Q/\sqrt{3})\{0,-1\}=-\sqrt{3}/3$, where $\Q=(2/3,-1/3)$. Clearly, this is incorrect since three quarks with electric charges $+2/3$ obviously form a composite baryon with an electric charge of $+2$ not $+2\sqrt{3}/3$. Similarly, three quarks with electric charges $-1/3$ form a composite baryon with an electric charge of $-1$ not $-\sqrt{3}/3$. Moreover, if we choose not to employ the proposed normalization factor $\sqrt{N}$, it is true that the projections of the vector $3\Q$ upon the 2-space coordinate axes $\{1,0\}$ and $\{0,-1\}$ do yield  correct electric charges $+2$ and $-1$ as required \cite{33}. However, in this case the baryon number of the composite is wrong, and there are no corresponding projection axes available for the \emph{intermediate} electric charges between $+2$ and $-1$ that can be formed by combining three quarks in various well known ways. In particular, there is no way to obtain electric charges $0$ and $+1$ by projecting the vector $3\Q$ upon the 2-space axes $\{1,0\}$ or $\{0,-1\}$.

Considerations of the foregoing kind lead to the conclusion that \emph{like} vector coordinates for two or more fundamental fermions \emph{cannot} be added or superimposed ($\Q$ with $\Q$, $\U$ with $\U$, $\V$ with $\V$), the way quantum states or state vectors can  be superimposed. Only the flavor-defining scalar \emph{charges} associated with the flavor-defining coordinates $\Q$, $\U$ and $\V$ for two or more fundamental fermions, can be combined or superimposed. That is, because flavor eigenstates are to be described in terms of various simultaneously-observable, global, flavor-defining (scalar) charge-like quantum numbers (associated with $\Q$, $\U$ or $\V$), it is simply \emph{not} physically meaningful to superimpose like coordinates $\Q$, $\U$ or $\V$ for two or more particles [8].

The foregoing discussion clearly demonstrates that the dynamical significance of the new flavor-defining fermion coordinates $\Q$, $\U$ and $\V$ is quite different than say, the intrinsic angular momentum or spin coordinates for two different fermions. Angular momentum vectors for two different particles can be combined, but the corresponding (like) $\Q$-, $\U$-, and $\V$-vector coordinates cannot. Nevertheless, the triad of 2-vector coordinates ($\Q$, $\U$ and $\V$) used to represent a flavor eigenstate, is expected to change or \emph{evolve} during certain kinds of \emph{transitions} or \emph{interactions}, simply because the wave function or state vector, which depends, in part, on flavor-defining global scalars associated with $\Q$, $\U$ and $\V$, necessarily, \emph{evolves}.

Clearly, the expectation that the vector triad \emph{evolves} could have experimental consequences. As we will demonstrate in a later section, the evolution of vector triads, as \emph{constrained} by new underlying physics, could lead to an explanation for, among other things, recent observations of solar neutrino mixing.

\subsection{Quantization of the coordinates U and V}

Because the fermion-number operator $\fop$ is a 2 by 2 matrix, its continuation (see Sec.\ 2.1.3), namely $\fv$, must likewise be a 2 by 2 matrix. Thus we have a natural explanation for why there must be \emph{two} new linearly-independent flavor-defining fermion coordinates $\U$ and $\V$ associated with quarks and leptons, and their hypothetical superpartners \cite{19}. Note that we have just \emph{two} new (2-vector) coordinates $\U$ and $\V$ because the resultant 2-space can maintain at most two linearly independent 2-vector resolutions of $\Q$ at a time \cite{31}. The question now arises as to how it is that these fermion coordinates can be \emph{quantized}.

Since we do not have a first-principles understanding of the proposed quantization in terms of underlying dynamics, we will appeal to simple heuristic arguments that serve not only to place the question in perspective, but also yield \emph{definite} rules for quantizing these coordinates.

\vbox{
\begin{center}
\begin{psfrags}
\includegraphics[width=4in]{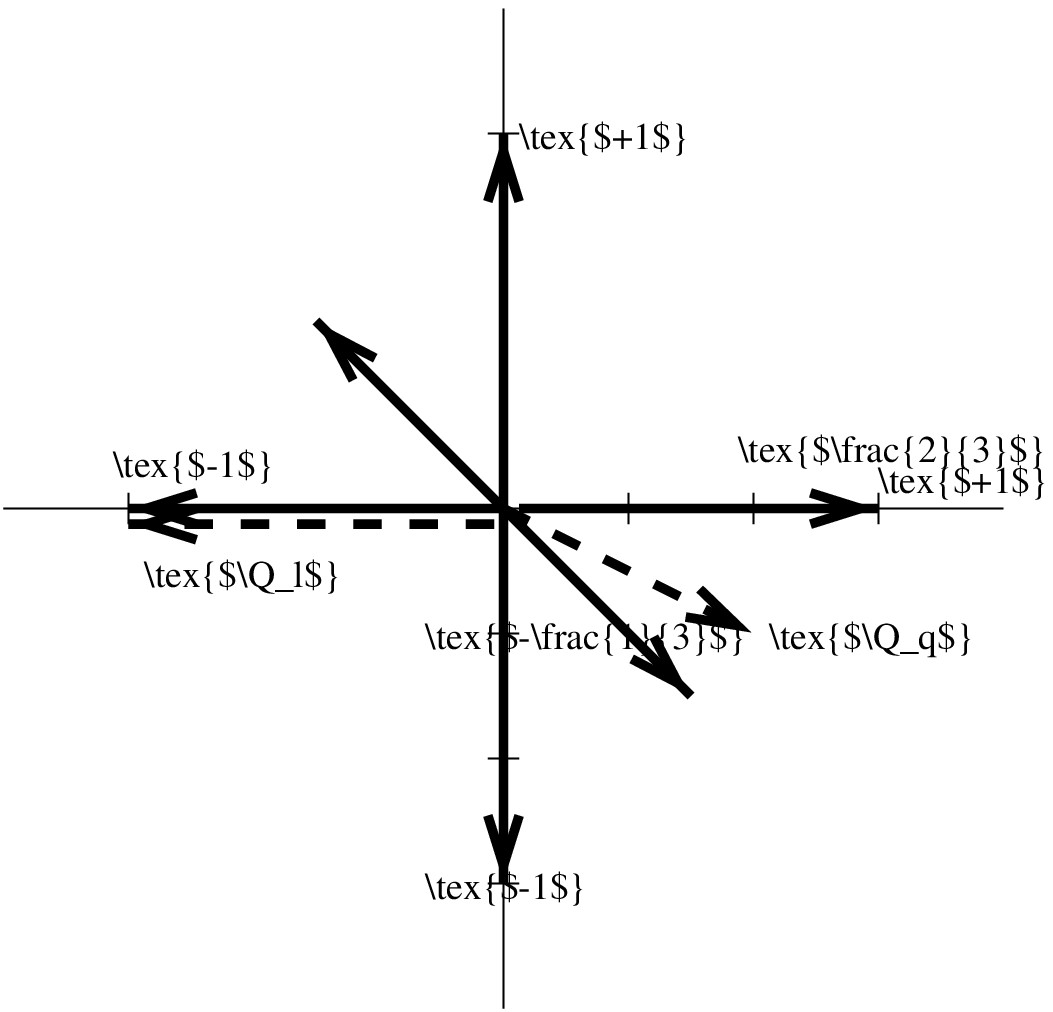}
\end{psfrags}
\end{center}
\begin{flushleft}
{\bf Figure 2}.
{This figure illustrates the \emph{six} $\U$-vector coordinates (solid
arrows),
and the \emph{two} ``electric charge'' or $\Q$-vector coordinates (dotted
arrows) associated with \emph{matter} particles (quarks $q$ and leptons $l$),
namely,
$\Q_q$ and $\Q_l$. The horizontal ``dotted'' arrow is offset from the origin
for
 purposes of illustration. It is assumed to pass through the origin.}
\end{flushleft}
}
\bigskip

\subsubsection{Quantization of the U-vector}

Write the 2-vectors $\U$ and $\V$ in terms of component charges, which we would like to quantize so as to quantize the coordinates $\U$ and $\V$. That is, write them as row $(\quad,\quad)$ or column $\{\quad,\quad\}$ vectors thus; $\U=(u_1,u_2)$ and $\V=(v_1,v_2)$ or $\U=\{u_1,u_2\}$ and $\V=\{v_1,v_2\}$, respectively. Now argue that \emph{it should take only one charge type} (\emph{i.e., $u$ or $v$}) \emph{to distinguish ``up'' and ``down'' type flavors} within any and all ``up''-``down'' type flavor doublets. If this is so, we can assume, without loss of generality, that this is the charge $u$ (i.e., $u_1$ or $u_2$). Then, since the charge $v$ does \emph{not} distinguish ``up'' and ``down'' type flavors, it must be true that $v_1=v_2$ for every flavor doublet of fundamental fermions. And, therefore, \emph{given the metric in} (11), \emph{the quantized charge $\V^2$ has but one value, namely, $\V^2=(v_1,v_2)\bullet\{v_1,v_2\}=(v^2_1-v^2_2)=0$ for all fundamental fermions}. That is, all fundamental fermions are ``neutral,'' i.e., they are \emph{singlets} with respect to the global charge $\V^2$.

Then, since $\Q=\U+\V$, we can write $\Q$ in column vector form in terms of $\U$ and $\V$ as $\Q=\{q_1,q_2\}=\{u_1,u_2\}+\{v_1,v_2\}$. Given that the difference between ``up'' and ``down'' \emph{electric} charges for \emph{all} fundamental fermions defines the fundamental (quantum) \emph{unit} of electric charge (see Eqs.\ 18 and 20), namely, $q_1-q_2=\pm 1$, and given that $v_1=v_2$ for all fundamental fermions, the vector $\U$ must have the column-vector form $\U=\pm\{a,a-1\}$ for \emph{all} fundamental fermions.

Now consider the nature of the charges $u$ (i.e., $u_1$ and/or $u_2$), which are the components of $\U=\{u_1,u_2\}$. Because the charge $u$ distinguishes ``up'' flavors from ``down'' flavors, and because it is a \emph{charge}, it must be possible for this charge, like \emph{all} known charges, to assume the value \emph{zero}. Then, given that $u$-charge can be zero in certain cases, and given the form $\U=\pm\{a,a-1\}$, we see immediately that the quantized values of $a$ must include the values $a=0$ or 1. This means that there are at least four $\U$-vectors, namely, the vectors $\U=\pm\{0,-1\}$ and $\U=\pm\{1,0\}$. And, given the metric (11), this means that the quantized charge $\U^2$ must include the values $\U^2=\pm 1$.

Finally, the quantized charge $\U^2$, like \emph{all} known charges, should be able to assume the value \emph{zero} for which $a=\frac{1}{2}$. Then the quantized values $a$ are found to be limited to the \emph{finite} sequence consisting of the \emph{three} values $a=0$, $\frac{1}{2}$, $1$. Similarly, we find that the quantized values of the charge $\U^2$ are found to be limited to the \emph{finite} sequence of \emph{three} values $\U^2=0$, $\pm 1$, while the associated quantized $\U$-vectors are limited to the \emph{finite} sequence of \emph{six} vectors (see Fig.\ 2), namely, $\U=\pm\{1/2, -1/2\}$, $\pm\{1,0\}$, $\pm\{0,-1\}$.

\subsubsection{Family replication and the number of families}

Clearly, if we limit the discussion to matter particles---the same arguments apply to antimatter particles and the vectors $\Q^c_q$ and $\Q^c_l$---we note that because the vector $\Q_q$ applies to quarks and $\Q_l$ applies to leptons, some of the $\U$- and $\V$-vectors (and associated global charges) will apply  to quarks, but \emph{not} to leptons, and vice versa. So, it is appropriate to ask which of the six possible $\U$-vectors go with the vector $\Q_q$, and which of the six possible $\U$-vectors go with the vector $\Q_l$. Once this question is answered, one can easily determine which $\V$-vectors go with the vector $\Q_q$, and which $\V$-vectors go with the vector $\Q_l$. In other words, we will be able to determine which $\U$-vectors go with which $\V$-vectors to form a vector triad $(\Q, \U, \V)$ that serves to describe \emph{flavor doublets} of quarks or leptons. This information  will tell us, in turn, how many \emph{generations} of quarks and leptons there are, and hence how many \emph{families} of fundamental fermions there are.

By inspection of Figure 2 we see that \emph{the requisite form for $\V$-vectors, namely}, $\V=\{v,v\}$ cannot \emph{be maintained if $\U$-vectors in quadrant II} (\emph{IV}) \emph{are associated with the vector $\Q_q$} ($\Q_l$) \emph{in quadrant IV} (\emph{II}). Therefore, we discover that the \emph{single} vector $\Q_q$ ($\Q_l$) in quadrant IV (II), can only be associated with the \emph{three} $\U$-vectors located in the \emph{same} quadrant. This simple result immediately tells us that, because there are only \emph{six} $\U$-vectors to choose from---three of which fall in quadrant IV, and three of which fall in quadrant II---there are only \emph{three} generations of quark flavor doublets and \emph{three} generations of lepton flavor doublets.

This is a remarkable result, for it means that a most unexpected explanation for why there are three families ``automatically'' emerges from the 2-space description. Of course, the underlying dynamics responsible for the 2-space, its quantization, and its associated selection rules  would, if available, presumably provide a much more detailed explanation \cite{29}.

\subsubsection{Quantization of the V-vector}

Given the four $\Q$-vectors (see Fig.\ 1), the six $\U$-vectors from Section 3.3.1 and Figure 2, the associations between vectors such as $\Q_q$, $\Q_l$ and the six $\U$-vectors (see Sec.\ 3.3.2), and the relation $\Q=\U+\V$, we have the following sequence of eleven quantized $\V$-vectors, consistent with the requisite form $\V=\{v,v\}$, namely, $\V=\{0,0\}$, $\pm\{1/6,1/6\}$, $\pm\{-1/3,-1/3\}$, $\pm\{-1/2,-1/2\}$, $\pm\{2/3,2/3\}$, $\pm\{-1,-1\}$ where the upper plus signs refer to matter, and the lower minus signs refer to antimatter. And, from the foregoing quantized $\U$- and $\V$-vectors we have the sequence of eleven quantized global charges $\U\bullet\V= 0$, $\pm 1/6$, $\pm1/3$, $\pm1/2$, $\pm2/3$, $\pm 1$.

Here, the $\V$-vectors $\pm\{1/6, 1/6\}$, $\pm\{-1/3, -1/3\}$ and $\pm\{2/3, 2/3\}$ apply to quarks (upper plus signs) and antiquarks (lower minus signs), while the $\V$-vectors $\{0,0\}$, $\pm\{-1/2,-1/2\}$ and $\pm\{-1,-1\}$ apply to leptons (upper plus signs) and antileptons (lower minus signs). Similarly, the charges $\U\bullet\V=\pm1/6, \mp 1/3, \pm 2/3$ apply to quarks (upper signs) and antiquarks (lower signs), while the charges $\U\bullet\V=0, \mp 1/2, \mp 1$ apply to leptons (upper signs) and antileptons (lower signs). Note that the scalar components of $\V$-vectors are proportional to the associated charges $\U\bullet\V$.

\subsubsection{A natural basis for family membership of flavor doublets}

The discussion in Section 3.3.2 makes it clear that there are only \emph{three} quark-lepton families. But, what determines which \emph{quark} flavor doublet goes with which \emph{lepton} flavor doublet to make a particular quark-lepton \emph{family}? In other words, how does one decide which quark $\U$-vector goes with which lepton $\U$-vector?

The question of family membership of flavor doublets is \emph{not} answered in the standard model of particle physics where these family assignments are made, more-or-less, on the basis of the mass hierarchy of the average family member. However, in the 2-space, there are other, more ``natural,'' ways of assigning flavor doublets to families (see Ref.\ 9 for a detailed discussion). We note first the fact that two of the three \emph{quark} flavor doublets associated with the three $\U$-vectors $\U=\{1/2, -1/2\}$, $\{1,0\}$ and $\{0,-1\}$, are ``equidistant'' from the ``intermediate'' $\U$-vector $\{1,0\}$, with respect to a matrix $\R$ given by
\begin{equation}
\R=\left(\begin{array}{rr}
0 & -2 \\
-1 & -1 \end{array}\right).
\end{equation}
That is, the matrix $\R$ is a Lorentz-invariant ``measure'' of the ``distance'' between certain $\U$-vectors, and their associated flavor doublets, thus \cite{34, 35}
\begin{equation}
\{1/2, -1/2\} \mathop{\longrightarrow}\limits^{\R} \{1,0\} \mathop{\longrightarrow}\limits^{\R} \{0,-1\}.
\end{equation}
Similarly, two of the three \emph{lepton} flavor doublets associated with the three $\U$-vectors $\{-1/2$,\break
 $1/2\}$, $\{-1,0\}$ and $\{0,1\}$, are also ``equidistant'' from the ``intermediate'' $\U$-vector $\{-1,0\}$, with respect to the \emph{same} matrix $\R$ since
\begin{equation}
\{-1/2, 1/2\} \mathop{\longrightarrow}\limits^{\R} \{-1,0\} \mathop{\longrightarrow}\limits^{\R} \{0,1\}.
\end{equation}
In other words, the $\U$-vector \emph{pairs} [$\{1/2, -1/2\}$ and $\{-1/2, 1/2\}$] or [$\{1, 0\}$ and $\{-1,0\}$]  or  [$\{0,-1\}$ and $\{0,1\}$] are at the \emph{same} ``distance'' or ``location'' with respect to the single ``distance'' measure $\R$. That is, they are associated with, or are ``in,'' the \emph{same} family. To state this differently, \emph{the $\U$-vector describing a quark flavor doublet in a given family is the additive \emph{inverse} of the $\U$-vector describing a lepton flavor doublet in the \emph{same} family, and vice versa}.

Because of the foregoing relations and identifications, it would seem very natural to assign the $\U$-vectors $\pm\{1/2, -1/2\}$ to the \emph{first} family, $\pm\{1,0\}$ to the \emph{second}, or intermediate family, and $\pm\{0,-1\}$ to the \emph{third} family, where the upper $+$ signs refer to \emph{quarks} and the lower $-$ signs refer to \emph{leptons}. While the assignment of the $\U$-vectors $\pm\{1,0\}$ to the \emph{second}, or intermediate, family seems unambiguous, what prevents us from assigning the $\U$-vectors $\pm\{1/2, -1/2\}$ to the \emph{third} family, and the $\U$-vectors $\pm\{0,-1\}$ to the \emph{first} family?

After all, we could just as well have used the multiplicative \emph{inverse} of the matrix $\R$, namely, $\R^{-1}$ where
\begin{equation}
\R^{-1}=\left(\begin{array}{rr}
\frac{1}{2} & -1 \\[2mm]
-\frac{1}{2} & 0 \end{array}\right),
\end{equation}
to be the Lorentz-invariant measure of ``distance,'' in which case we would have had the following relations and associations
\begin{equation}
\pm\{1/2, -1/2\} \mathop{\longleftarrow}\limits^{\R^{-1}} \pm\{1,0\} \mathop{\longleftarrow}\limits^{\R^{-1}} \pm\{0, -1\}.
\end{equation}
Clearly, from these relations and associations it would seem just as natural to assign the $\U$-vectors $\pm\{1/2, -1/2\}$ to the \emph{third} family, $\pm\{1,0\}$ to the \emph{second}, or intermediate family, and $\pm\{0, -1\}$ to the \emph{first} family. How then, are we to decide if the $\U$-vectors $\pm\{1/2, -1/2\}$, are to be assigned to the \emph{first} or the \emph{third} family?  The 2-space description of quarks and leptons provides a simple ``answer'' to this question as well.

It turns out (see Ref.\ 9, pp.\ 56--58) that the vector triads $(\Q, \U, \V)$ associated with both quarks and leptons have the \emph{highest} possible \emph{degree} of symmetry with respect to the matrix transformation $\fv$, \emph{only} when the associated $\U$-vectors are $\pm\{1/2, -1/2\}$. And, because the ``ground state'' is typically the most probable state, with the highest degree of symmetry into which other less symmetrical states tend to \emph{evolve} \cite{36}, it is natural to assume that the $\U$-vectors $\pm\{1/2, -1/2\}$ are associated with the \emph{first} or ``ground state'' family. Recall that the members of the first family have the lowest average mass of any family members, and so the first family naturally resembles a kind of ``ground state.''

The foregoing associations and identifications, together with the specific family assignments of the \emph{experimentally} observed quarks and leptons, are elaborated elsewhere (see Ref.\ 9). However, it should be clear---as we have repeatedly emphasized throughout this paper---that the foregoing heuristic 2-space relations and associations must be a consequence of some ``deeper,'' more fundamental, ``layer'' of physics that is responsible for the 2-space, its quantization, and its associated selection rules.

\subsection{The nature of transitions of vector triads}

Because strong and electromagnetic interactions have no effect on flavor indices, they can have no effect on the flavor-defining fermion coordinates. Hence, vector triads $(\Q, \U, \V)$ will be unaffected by \emph{strong} and \emph{electromagnetic} interactions. However, in general, this is not the case for \emph{weak} interactions.

Because the quantum state of an ``isolated'' fundamental fermion such as a quark (eventually) evolves via weak interactions, so must the associated vector triad, consisting of the three 2-vectors ($\Q$, $\U$ and $\V$) \emph{evolve}, and vice versa. And, if we knew the dynamics \cite{29} underlying the 2-space, presumably we would be in a position to calculate, from first principles, such things as Cabibbo-Kobayashi-Maskawa (CKM) matrix elements. However, as this dynamics is currently unknown, we will have to satisfy ourselves here with a less ambitious description based on mathematical properties of the 2-space, and certain experimental facts about transitions of fundamental fermions. Consider first the so-called ``diagonal'' weak transitions of quarks.

Diagonal weak transitions of quarks are ``up''-``down'' or ``down''-``up'' type transitions involving the two quark flavors (flavor eigenstates) ``located'' within the \emph{same} flavor doublet. From experimental data on CKM matrix elements, these diagonal transitions are known to be far more probable than the so-called ``off-diagonal'' transitions, which are relatively suppressed. What do these experimental facts teach us about the \emph{evolution} of vector triads?

In the case of diagonal transitions, it is clear that the vector triad associated with the \emph{initial} quark state, and the vector triad associated with the \emph{final} quark state, are identical, i.e., they are \emph{congruent}. By comparison, the initial and final state triads associated with all ``off-diagonal'' weak transitions of quarks are \emph{different}. These observations suggest that the underlying dynamics tends to ``prefer'' transitions in which the associated vector triad is \emph{unchanged}. Clearly, such transitions would be characterized by the ``selection rules'' (use $\Q=\U+\V$)
\begin{equation}
\Delta\Q=\Delta\U=\Delta\V=0.
\end{equation}
Moreover, since $\Q$, $\U$ and $\V$ don't change in such transitions, we also have the selection rules
\begin{equation}
\Delta\Q^2=\Delta\U^2=\Delta\U\bullet\V=0,
\end{equation}
which means that the baryon number $B=\Q^2$, the charge $\U^2$, and the charge $\U\bullet\V$, are separately \emph{maintained} or \emph{conserved} in such transitions \cite{37}.

While quark masses play a significant role in determining the magnitudes of CKM matrix elements, there appears to be a trend here, namely, this: ``Easy'' transitions (i.e., diagonal transitions \emph{within} a quark generation) are those for which the associated vector triad does \emph{not} change, and ``hard'' transitions (i.e., off-diagonal transitions \emph{between} different quark generations) are those for which it \emph{does} change.

To be more specific, while there are some 24 flavor and antiflavor eigenstates of quarks and leptons \cite{17}, there are just 12 \emph{distinct} vector triads. That is,  the vector triad associated with each flavor doublet is associated with each  of \emph{two} flavor or antiflavor eigenstates. This means that all diagonal transitions involve the \emph{same} vector triad, and all off-diagonal transitions involve \emph{different} (distinguishable) vector triads.

What is it that differs from vector triad to vector triad? Of course, charge-like quantum numbers such as
\begin{equation}
\Q^2, \U^2, \mbox{ and } \U\bullet\V
\end{equation}
can sometimes differ. But there are other, more subtle, things that can differ as well. It happens that different vector triads, and certain of their associated 2-vector coordinates, are ``isolated'' from one another in a \emph{topological} sense. Take for example the four 2-vectors $\Q_q$, $\Q^c_q$, $\Q_l$, and $\Q^c_l$ illustrated in Figure 1. 

Because there are no \emph{continuous} transformations (associated with the metric in Eq.\ 11) available to convert the 2-vectors $\Q_q$, $\Q^c_q$, $\Q_l$ and $\Q^c_l$ one into the other, these four $\Q$-vectors are, necessarily, \emph{topologically distinct}. It is certainly conceivable that \emph{this fact could be an important underlying reason for the separate conservation of baryon number $B$ and lepton number} $L$ in most physical processes. In particular, because the $\Q$-vectors are \emph{topologically distinct}, underlying dynamics could conceivably dictate that there are topological energy ``barriers'' that \emph{maintain} these topological distinctions, leading to the 2-vector selection rule \cite{39}
\begin{equation}
\Delta\Q=0,
\end{equation}
and consequently to the selection rule (Note that $\Q^2=B$ or $L$)
\begin{equation}
\Delta\Q^2=0,
\end{equation}
which should then apply to \emph{all} fundamental fermions \cite{39}.

Of course, the  discussion in the previous paragraph involves only \emph{one} of the three 2-vector coordinates (i.e., $\Q$) associated with a vector triad $(\Q, \U, \V)$. There are other topological distinctions that can be drawn, which involve all \emph{three} 2-vectors associated with a particular vector triad. For example, as shown elsewhere (see Ref.\ 9, pp.\ 56, 57, and Appendix A in Ref.\ 13), and as elaborated in the following section, there are also topological distinctions based on how an \emph{entire} vector triad $(\Q,\U, \V)$ \emph{transforms} under the matrix $\fv$. It happens that some vector triads transform like M\"obius
strips, and some transform like cylinders under $\fv$. Clearly, if there were underlying topological energy ``barriers'' that act to inhibit changes in the topology of  \emph{entire} vector triads, this situation could have experimental consequences \cite{38}. And, if experiments were to eventually confirm that such topological energy ``barriers'' exist, a new level of dynamics ``below'' or ``beyond'' the standard model of particle physics would seem to be required to explain them.

\subsection{Does the dynamics underlying the 2-space impose topological constraints on 3-flavor neutrino mixtures?}

As alluded to in the previous section, it is important to understand that the new 2-space description of fundamental fermions (quarks and leptons) provides a distinction between these particles that goes beyond differences that can be explained by mass differences alone. For example, in the standard model of particle physics the only difference between the $u$, $c$ and $t$ quarks is that they have different masses. Otherwise, these particles experience identical strong and electroweak interactions. Moreover, the separate conservation (in strong and electromagnetic interactions) of quantum numbers such as ``charm'' and ``truth'' can be attributed to certain unavoidable ``accidental symmetries'' associated with the (renormalizable) Lagrangian describing the  interactions of these particles \cite{7}.

Taken at face value, these accidental symmetries would seem to imply that there are no internal ``wheels and gears'' that would distinguish a $u$ quark from a $c$ quark, for example. But, if the string theories are correct, these particles would be associated with different ``handles'' on the compactified space (see Ref.\
4, Vol.\ 2, p.\ 408), and so would be different in this \emph{additional} sense. Likewise, in the present non-Euclidean
2-space description, \emph{topological} differences in addition to a variety of (global) 2-scalars, which are  \emph{indirectly} related to the  accidental symmetries of the Lagrangian, serve to provide further distinctions between matter particles.

A possible experimental signal of such internal differences is to be found in the recent observations at the Super Kamiokande and SNO of (nearly) bi-maximal $\nu_\mu-\nu_\tau$ neutrino mixing \cite{43}. Models which begin by positing a neutrino mass-matrix, and associated mixing-parameters, such as the three-generation model proposed by Georgi and Glashow
\cite{40}, do an acceptable job of describing the observations. However, (nearly) bi-maximal $\nu_\mu-\nu_\tau$ mixing may have a deeper explanation in terms of internal topological differences (in the non-Euclidean 2-space) between $\nu_e$, and $\nu_\mu$ or $\nu_\tau$ neutrinos.

With respect to the internal transformation $\fv$, the topology of the non-Euclidean ``vector triad'' (see Ref.\ 9, p.\ 57,  and the qualifying remarks in Appendix A of Ref. 13) associated with the $\nu_e$ ($\nu_\mu$ or $\nu_\tau$), is found to be that of a cylinder (M\"obius strip). And, \emph{assuming that a change in topology} (\emph{of vector triads}) \emph{during neutrino mixing is suppressed by energy ``barriers,'' or other topological ``barriers''} described by physics ``beyond'' the standard model, while neutrino mixing without topology-change is (relatively) enhanced \cite{38, 41, 42}, one can readily explain the experimental observation of (nearly) bi-maximal $\nu_\mu-\nu_\tau$ neutrino mixing---at least (nearly) bi-maximal $\nu_\mu-\nu_\tau$ mixing over long distances, where the proposed topological influences are expected to be \emph{cumulative}. If this qualitative explanation (see detailed calculations below) is basically correct, then it follows that the neutrino mass-matrix, and associated 
mixing-parameters needed to explain (nearly) bi-maximal $\nu_\mu-\nu_\tau$ neutrino mixing, would be the \emph{result} of new physics (i.e., dynamics) associated with these deeper (internal) topological differences between neutrinos, and not their \emph{cause}.

\subsubsection{Conventional description of 3-flavor neutrino mixing}

Except where explicitly prevented by some ``absolute'' conservation law
(e.g., the conservation of electric charge or spin angular momentum),
quantum mechanics generally permits transitions between states having
\emph{different} topologies \cite{41}. While a change in topology may be energetically
(or otherwise) inhibited by underlying dynamics, unavoidable quantum fluctuations are expected to
\emph{catalyze} such processes. Hence, there is always the possibility of
\emph{mixing} between otherwise similar quantum states (e.g., $\nu_e$, $\nu_\mu$ and $\nu_\tau$) associated with vector triads having \emph{distinct}
topologies \cite{42}.

As described previously---with respect to the matrix transformation $\fv$---the topology associated with both the 
$\nu_\mu$  and $\nu_\tau$
is found to be that of a M\"obius strip (Ref.\ 47, p.\ 143).  By contrast,  
the topology associated with the $\nu_e$
[with respect to $\fv$] is that of a cylinder.
And because it is reasonable to \emph{assume} that a change in topology 
during transitions tends to be \emph{suppressed} \cite{38},  the foregoing topological distinctions between 
neutrinos may help explain recent  observations of (nearly) 
bi-maximal $\nu_\mu$-$\nu_\tau$  
mixing  \cite{43}. However, it is important to stress at the outset that there are good reasons for believing that similar topological distinctions (alone) among quarks, while also present, do \emph{not} play an important role in $d$, $s$, $b$ quark mixing \cite{44}.

At ``birth'' via weak decays, or upon detection via weak capture interactions, neutrinos have a \emph{definite} flavor and associated topology \cite{45}. However, between birth and detection they are in a mixed state having no definite flavor or associated topology. In this intermediate region the probability of flavor (or topology) maintenance and/or change \emph{oscillates}. Only at great distances from the neutrino source do these oscillations finally ``damp out.''

In the conventional description of three-flavor neutrino mixing \emph{flavor eigenstates} are related to neutrino \emph{mass eigenstates} (states of definite mass) via a unitary 
CKM-like ``mixing'' matrix $U_{\alpha i}$, as follows (Ref.\ 3, p.\ 365 and Ref.\ 46)
\begin{equation}
\nu_\alpha=\sum^3_{i=1}U_{\alpha i}\nu_i.
\end{equation}
Here, $\nu_i=\nu_1, \nu_2$ and $\nu_3$ are mass eigenstates with mass eigenvalues $m_1, m_2$ and $m_3$, respectively, while $\nu_\alpha=\nu_e$, $\nu_\mu$ and $\nu_\tau$ are flavor eigenstates. Using a conventional parameterization
for $U_{\alpha i}$, (61) becomes
\begin{equation}
\left(\begin{array}{c}
\nu_e\\
\nu_\mu\\
\nu_\tau\end{array}\right) =
\left(\begin{array}{ccc}
c_1 & s_1c_3, & s_1s_3\\
-s_1c_2,&c_1c_2c_3-s_2s_3e^{i\delta}, & c_1c_2s_3+s_2c_3e^{i\delta}\\
-s_1s_2, & c_1s_2c_3+c_2s_3e^{i\delta}, & c_1s_2s_3-c_2c_3e^{i\delta}\end{array}\right)\;
\left(\begin{array}{c}
\nu_1\\
\nu_2\\
\nu_3\end{array}\right),
\end{equation}
where $c_i\equiv\cos \theta_i$ and $s_i\equiv\sin\theta_i$, and $\delta$ is associated with a Dirac-type CP-noninvariant phase factor
$e^{i\delta}$.

Using (62) it can be shown that the probability of detecting a neutrino of flavor type $\beta$ at a distance $X$ from a source of neutrinos of flavor type $\alpha$ is given by \cite{46}
\begin{equation}
P_{\nu_\alpha\to\nu_\beta} = \sum^3_{i=1}|U_{\alpha i}|^2|U_{\beta i}|^2 + \sum^3_{i\ne j} U_{\alpha i}U^\ast_{\beta i} 
U^\ast_{\alpha j}U_{\beta j}\cos\left(\frac{2\pi X}{l_{ij}}\right).
\end{equation}
Here, the so-called \emph{oscillation lengths} $l_{ij}$ are given by $l_{ij}=2\pi/(E_i-E_j)$, where the total relativistic
energy differences in a beam of neutrinos having \emph{fixed} momentum $p$ are $E_i-E_j=(m^2_i-m^2_j)/2p$.

Examination of (63) shows that at the neutrino source ($X=0$, and $t=0$) the probability $P_{\nu_\alpha\to\nu_\beta}$ reduces, as expected, to the $3\times 3$ identity matrix 
\begin{equation}
P_{\nu_\alpha\to\nu_\beta}\Bigg|_{X=0}=I_3,
\end{equation}
while at ``intermediate'' distances from the neutrino source $(X\approx l_{ij})$, the probability $P_{\nu_\alpha\to\nu_\beta}$ undergoes \emph{oscillations}.  Finally, at great distances from the neutrino source $(X\gg l_{ij})$, all time- or distance-dependent oscillations ``damp out,'' and we are left with a $3\times 3$ matrix of time-average probabilities \cite{46}, namely, 
\begin{equation}
\langle P_{\nu_\alpha\to\nu_\beta}\rangle = \sum^3_{i=1} |U_{\alpha i}|^2|U_{\beta i}|^2.
\end{equation}
From (65) it is clear that this matrix ($\alpha=$ row index, $\beta=$ column index), which describes long-distance neutrino mixtures, is \emph{symmetric}. Keeping in mind this symmetry, and calling this matrix $M$, one has
\begin{equation}
M = \left(\begin{array}{ccc}
\langle P_{\nu_e\to\nu_e}\rangle, & \langle P_{\nu_e\to\nu_\mu}\rangle, & \langle P_{\nu_e\to\nu_\tau}\rangle \\
\langle P_{\nu_\mu\to\nu_e}\rangle, & \langle P_{\nu_\mu\to\nu_\mu}\rangle, & \langle P_{\nu_\mu\to\nu_\tau}\rangle \\
\langle P_{\nu_\tau\to \nu_e}\rangle, & \langle P_{\nu_\tau\to\nu_\mu}\rangle, & \langle P_{\nu_\tau\to\nu_\tau}\rangle\end{array}\right).
\end{equation}
Note that all rows and columns of $M$ must sum to unity (total probability 1).

Using $M$ we can describe the expected neutrino flavor content at a great distance from a neutrino source (e.g., the sun or a supernova) as follows \cite{40}
\begin{equation}
\{D_e, D_\mu, D_\tau\}=M\{B_e, B_\mu, B_\tau\},
\end{equation}
where $\{\quad\}$ signifies column vectors, and $D_\alpha$ is the number of \emph{detected} neutrinos 
of definite flavor $\nu_\alpha$, and $B_\alpha$ their number at ``birth'' at some distant neutrino source. Note that because neutrinos are assumed to be conserved, the total number of neutrinos at birth equals their number upon ``detection,'' namely,
\begin{equation}
B_e+B_\mu+B_\tau=D_e+D_\mu+D_\tau.
\end{equation}

\subsubsection{Proposed topological constraints on 3-flavor neutrino mixing}

Given that the $\nu_e$ ($\nu_\mu$ or $\nu_\tau$) neutrino flavor
has the associated topology of a cylinder
(M\"obius strip) with respect to the internal transformation $\fv$,
and assuming that topological constraints are the
\emph{primary} determinants of the matrix $M$ describing long-distance neutrino 
mixtures, the matrix $M$ is easily determined.
 To accomplish this we need only apply the
following very general principle to neutrino-neutrino transitions:
\medskip

\emph{All other things being equal, any neutrino flavor $\nu_\alpha$ {\rm(}i.e.,
$\nu_e$, $\nu_\mu$ or $\nu_\tau${\rm)}, which undergoes neutrino-neutrino
transitions that change the associated  neutrino topology, will tend to be suppressed,
while neutrino-neutrino transitions that maintain the associated neutrino topology will
tend to be {\rm(}relatively{\rm)} enhanced} \cite{38}.
\medskip

To this principle we add the following corollary,
\medskip

\emph{All other things being equal, because the $\nu_\mu$ and $\nu_\tau$ neutrinos have
the same associated topology, they will act the same way in all neutrino-neutrino
transitions {\rm (}involving long-distance neutrino mixtures{\rm )}.}
\medskip

Given these principles,  and assuming as stated previously that topological constraints are the \emph{primary} determinants of the matrix $M$, we immediately have the following  \emph{topological} constraints on
long-distance neutrino mixtures:
\medskip

A. No matter what neutrino flavor $(\nu_\alpha)$ and associated topology one starts with at
some distant source (say the sun or a supernova), by the time the neutrino 
mixture reaches its
``equilibrium'' state (where all time-dependent oscillations have ``damped out''), it should contain \emph{equal} fractions of $\nu_\mu$
and $\nu_\tau$, because these neutrinos have the \emph{same} associated topology.
\medskip

B. Because the $\nu_\mu$ and $\nu_\tau$ neutrinos have the \emph{same} associated topology, if one
starts out with \emph{either} a pure $\nu_\mu$ \emph{or} a pure $\nu_\tau$
source, one should end up with the \emph{same} long-distance equilibrium
mixture of $\nu_e$, $\nu_\mu$ and $\nu_\tau$.
\medskip

C. If topology is the controlling factor in describing long-distance neutrino mixtures, then \emph{there should be absolutely no {\rm(}effective{\rm)} difference between the two functions {\rm(}of $U$-matrix mixing parameters}), which describe $\lan P_{\nu_e\to\nu_\mu}\ran$ \emph{and} $\lan P_{\nu_e\to\nu_\tau}\ran$, because the $\nu_\mu$ and $\nu_\tau$ neutrinos are associated with the \emph{same} topology.  Very loosely speaking we are assuming that these mathematical functions are
effectively ``topological invariants'' with respect to the exchange of flavor indices $\mu$ and $\tau$ (see Ref.\ 47, pp.\ 20 and 21). That is, not only are the two functions $\lan P_{\nu_e\to\nu_\mu}\ran$ and $\lan P_{\nu_e\to\nu_\tau}\ran$ required to be \emph{equal}, but they are also required to be equal, \emph{term-by-term} (i.e., they are required to be \emph{term-wise} equal). Similarly the three functions $\lan P_{\nu_\mu\to\nu_\mu}\ran$, $\lan P_{\nu_\mu\to\nu_\tau}\ran$ and $\lan P_{\nu_\tau\to\nu_\tau}\ran$ are required to be term-wise equal. 
\medskip

Constraints A), B) and C), together with Eqs.\ (67) and (68), dictate that the symmetric
matrix $M$ describing long-distance neutrino mixtures must have the
form expressed by
\begin{equation}
\left(\begin{array}{c}
D_e\\
D_\mu\\
D_\tau\end{array}\right) = \left(
\begin{array}{ccc}
a&b&b\\
b&c&c\\
b&c&c\end{array}\right) 
\left(\begin{array}{c}
B_e\\
B_\mu\\
B_\tau\end{array}\right).
\end{equation}

\subsubsection{Determination of the matrix $M$}
When the proposed topological constraints of Section 3.5.2 are applied to the conventional description of neutrino mixing (62), the matrix $M$ in (66), (67) and (69), is \emph{uniquely} determined. To see how this happens consider the following time-average probabilities (see Appendix B in Ref. 13 for details)
\begin{equation}
\langle P_{\nu_e\to\nu_\mu}\rangle = 2c^2_1 s^2_1 c^2_2 + 2s^2_1 s^2_3 c^2_3 (s^2_2 - c^2_1 c^2_2) + 2s^2_1 s_2s_3 c_1c_2c_3\cos \delta(s^2_3-c^2_3),
\end{equation}
and
\begin{equation}
\langle P_{\nu_e\to\nu_\tau}\rangle = 2 c^2_1 s^2_1 s^2_2 + 2s^2_1 s^2_3 c^2_3 (c^2_2 - c^2_1 s^2_2) - 2s^2_1 s_2 s_3 c_1 c_2 c_3 \cos\delta (s^2_3 - c^2_3).
\end{equation}
According to the proposed topological constraints of Section 3.5.2, these two  time-average probabilities must be 
\emph{term-wise equal}, and \emph{nonzero}. These requirements place \emph{three} constraints on the $U$-matrix mixing parameters, namely, $s^2_1>0$, $s^2_2=c^2_2=\frac{1}{2}$, and
\begin{equation}
s^2_1 s_2 s_3 c_1 c_2 c_3\cos\delta (s^2_3-c^2_3)=0.
\end{equation}
Note that (70) and (71) are term-wise equal as required, even with the minus sign preceeding the last term of (71) because this term is of zero magnitude.

Next, the topological constraint of Section 3.5.2, namely, $\langle P_{\nu_\mu\to\nu_\mu}\rangle = \langle P_{\nu_\tau\to\nu_\tau}\rangle$ can be realized provided the $U$-matrix
mixing parameters are further constrained by $s^2_3=c^2_3=\frac{1}{2}$, which also happens to satisfy (72). That is, given $s^2_2=c^2_2$ and $s^2_3=c^2_3$ one has (see Appendix B in Ref.\ 13 for details)
\begin{equation}
\langle P_{\nu_\mu\to\nu_\mu}\rangle = s^4_1 c^4_2 + 2c^4_2 c^4_3(c^2_1+1)^2 + 8c^2_1 c^4_2 c^4_3 \cos^2\delta,
\end{equation}
and
\begin{equation}
\langle P_{\nu_\tau\to\nu_\tau}\rangle = s^4_1 c^4_2 + 2c^4_2 c^4_3(c^2_1+1)^2 + 8c^2_1 c^4_2 c^4_3 \cos^2\delta.
\end{equation}

Now the topological constraints of Section 3.5.2 also require that $\langle P_{\nu_\mu\to\nu_\tau}\rangle$ be equal to (73) and (74). This places additional constraints on $M$, and the $U$-matrix mixing parameters.
Comparing (73) and (74) with the following expression from Appendix B in Ref.\ 13 for $\langle P_{\nu_\mu\to\nu_\tau}\rangle$, namely,
\begin{equation}
\langle P_{\nu_\mu\to\nu_\tau}\rangle = s^4_1 c^4_2 + 2c^4_2 c^4_3(c^2_1+1)^2 - 8c^2_1 c^4_2 c^4_3 \cos^2\delta,
\end{equation}
we see that the requisite equality $\langle P_{\nu_\mu\to\nu_\mu}\rangle = \langle P_{\nu_\tau\to\nu_\tau}\rangle = \langle P_{\nu_\mu\to\nu_\tau}\rangle$ leads to the constraint (note that $c^4_2=c^4_3>0$)
\begin{equation}
c^2_1 \cos^2\delta = 0.
\end{equation}
Assuming \cite{40} that $\sin^2\delta\ne 1$ or $\cos^2\delta\ne 0$ (i.e., CP violation is \emph{not} maximal),  
 we further determine from (76) that
\begin{equation}
c^2_1=0\mbox{ and }s^2_1=1.
\end{equation}
Note that (73), (74) and (75) exhibit the requisite term-wise equality [see item C) in Sec.\ 3.5.2] even with the minus sign preceeding the third term of (75), because this term is of zero magnitude.

Gathering together the predicted mixing-parameter constraints, namely, $c^2_2 = s^2_2$, $c^2_3 = s^2_3$, $s^2_1=1$ and $c^2_1=0$, we can express six of the nine time-average probabilities associated with the (symmetric) matrix $M$ as follows (see Ref.\ 13).
\begin{equation}
\langle P_{\nu_e\to\nu_e}\rangle  = 1-2c^4_3, 
\end{equation}
\begin{equation}
\langle P_{\nu_e\to\nu_\mu}\rangle  =  \langle P_{\nu_e\to\nu_\tau}\rangle = 2c^2_2 c^4_3,  
\end{equation}
and
\begin{equation}
\langle P_{\nu_\mu\to\nu_\mu}\rangle  = \langle P_{\nu_\tau\to\nu_\tau}\rangle = \langle P_{\nu_\mu\to\nu_\tau}\rangle = 
c^4_2(1+2c^4_3). 
\end{equation}
But, the previous arguments have shown that the $U$-matrix mixing parameters satisfy $c^2_2=c^2_3 = \frac{1}{2}$, which leads to the following specific numerical \emph{predictions}
\begin{equation}
\langle P_{\nu_e\to\nu_e}\rangle  = \frac{1}{2} = a, 
\end{equation}
\begin{equation}
\langle P_{\nu_e\to\nu_\mu}\rangle  = \langle P_{\nu_e\to\nu_\tau}\rangle = \frac{1}{4} = b,  \end{equation}
and
\begin{equation}
\langle P_{\nu_\mu\to\nu_\mu}\rangle  = \langle P_{\nu_\tau\to\nu_\tau}\rangle = \langle P_{\nu_\mu\to\nu_\tau}\rangle = \frac{3}{8} = c. 
\end{equation}
Given that $M$ is symmetric, note that (81), (82) and (83) are \emph{the} only \emph{matrix elements, which could be consistent with the proposed topological constraints of Section 3.5.2, and the requirement that $\sin^2\delta\ne 1$ or  $\cos^2\delta\ne 0$}.

Employing Eqs. (69) and (81--83), one finally has the \emph{prediction} \cite{40, 43}
\begin{equation}
M = \frac{1}{8} \left( \begin{array}{ccc}
4 & 2 & 2 \\
2 & 3 & 3 \\
2 & 3 & 3 \end{array}\right). 
\end{equation}
To summarize, \emph{the proposed} (\emph{qualitative}) \emph{topological constraints on the matrix $M$ {\rm (}see Sec.\ 3.5.2{\rm )} result in quantitative constraints on the $U$-matrix  mixing parameters {\rm (}see Eq.\ 62{\rm )}, namely,} ($s^2_2=c^2_2=c^2_3 = s^2_3 = \frac{1}{2}, s^2_1=1$ \emph{or} $c^2_1=0$ \emph{with} $\sin^2\delta\ne 1$ or $\cos^2\delta\ne 0$ assumed), \emph{which result, in turn,  in a} unique \emph{quantitative determination of the matrix} $M$ (see Eq.\ 84).

\subsubsection{Topology-maintaining and topology-changing influences in equilibrium?}

Not only is the matrix $M$ in (84) a \emph{unique} solution to the proposed topological constraints of Section 3.5.2 with $\sin^2\delta\ne 1$ or $\cos^2\delta\ne 0$, but it also has some very special properties that may eventually help reveal the deeper
\emph{dynamical} significance of this matrix. Note that (66), (69), and (84) have the very special property 
\begin{equation}
\langle P_{\nu_e\to\nu_e}\rangle = \langle P_{\nu_e\to\nu_\mu}\rangle + \langle P_{\nu_e\to\nu_\tau}\rangle.
\end{equation}
This equation says that the time-average probability that the associated $\nu_e$ topology  \emph{doesn't} change, namely,
\begin{equation}
P_{NC}=\langle P_{\nu_e\to\nu_e}\rangle = a, 
\end{equation}
and the time-average probability that the associated $\nu_e$ topology \emph{does} change, namely,
\begin{equation}
P_C=\langle P_{\nu_e\to\nu_\mu}\rangle + \langle P_{\nu_e\to\nu_\tau}\rangle = (1-a), 
\end{equation}
are \emph{equal}, namely,
\begin{equation}
P_C=P_{NC}.
\end{equation}

Now this equality looks very much like an ``equilibrium'' condition between those underlying physical influences that would act to \emph{change} the associated $\nu_e$ topology (quantum fluctuations), and those underlying physical influences that would act to \emph{maintain} the associated $\nu_e$ topology (e.g., topological energy ``barriers'').

We will now provide further support for this proposal. In particular, consider the ``joint'' probability \cite{48}
\begin{equation}
P=P_C\cdot P_{NC},
\end{equation}
and notice that
\begin{equation}
\frac{d P}{d a} = P_C\;\frac{d P_{NC}}{da} + P_{NC}\;\frac{dP_C}{d a}.
\end{equation}
From (86) and (87) this last equation reduces to
\begin{equation}
\frac{dP}{da} = P_C-P_{NC}.
\end{equation}
And, taking the second derivative, we also find
\begin{equation}
\frac{d^2P}{d a^2} = -2<0.
\end{equation}

Therefore, when $P_C=P_{NC}$, (91) and (92) tell us that the joint probability $P=P_C\cdot P_{NC}$ is a \emph{maximum}, namely, it characterizes some \emph{most probable} condition or ``state.'' And, this of course is the very essence of an ``equilibrium'' condition \cite{12}.
However, it must be understood that this hypothetical (long-distance) ``equilibrium'' between topology-changing, and topology-maintaining physical influences, is only a (cumulative) \emph{result} of \emph{deeper}, and largely unknown (short-distance), dynamical  processes in the vacuum, which first begin to
\emph{act on neutrinos at their source---thereby, eventually establishing the equilibrium condition $P_C=P_{NC}$---on time scales very much shorter than the time it takes for the time-dependent oscillations in neutrino mixtures to ``damp out.''}  This is an essential requirement if these 
hypothetical short-distance processes are to be responsible for ``selecting'' the (constant) $U$-matrix mixing parameters \emph{prior} to neutrino mixing.

\subsubsection{The current experimental situation in regard to 3-flavor neutrino mixtures}

The most recent analysis \cite{49} of Super Kamiokande and SNO data on solar neutrinos \cite{43} is in good agreement with the prediction given by (84). If this prediction is eventually verified by further detailed observations 
of neutrinos from distant astronomical sources (e.g., the sun or a supernova), and/or in long-baseline terrestrial experiments (e.g., see a description of proposed Kamland experiments in Ref.\ 50),
this will provide qualitative
support for the new description of fundamental fermions, which 
requires, among other things, that
the $\nu_e$ and ($\nu_\mu$  or $\nu_\tau$) neutrino flavors start, and end, ``life''
as \emph{topologically distinct} quantum objects (see Appendix A in Ref.\ 13, and Refs.\ 9, 38, 47). However, while it is conceivable that experiments could still \emph{falsify} (84), together with  the assumption \cite{40} that CP violation is \emph{not} maximal (i.e., $\sin^2\delta\ne 1$ or $\cos^2\delta\ne 0$), experimental verification of these things would \emph{not} confirm  the new description or the proposed \emph{topological} constraints of Section 3.5.2 \cite{51}. A much better theoretical understanding of the dynamical significance  of the new 2-space description of quarks and leptons, in terms of the new internal 2-vector coordinates, will be required before one can be confident that, if finally verified, (84) really is a reflection of new physics, e.g., underlying topological constraints \cite{51}. Clearly, additional (crucial) experimental consequences of the 2-space must be found if it is to be considered a serious candidate for a description of nature.

\subsection{Does the 2-space imply the Standard Model?}

Perhaps the best evidence we have for a ``layer'' of new physics---located somewhere between the unification scale, and the region of applicability of the standard model of particle physics---is simply that by constraining the possibilities, the 2-space description seems to \emph{imply} something very similar to the standard model. Consider the following facts in support of this hypothesis.

First, and foremost, we have found that owing to the 2-space parameter $v=\ln M_c$, where $M_c$ is a measure of the strong-color ``multiplicity,'' a symmetry associated with strong interactions such as $SU(3)_\color$ seems to be more-or-less implicit in the 2-space description. Second, because the 2-space determines that there are some 48 fundamental fermions and antifermions \cite{17}, together with their associated global charges---identified with the global charges associated with the accidental symmetries of the strong-electroweak Lagrangian---the form of the Lagrangian is severely restricted. Third, it is clear that the 2-space dictates that we are dealing here with \emph{flavor doublets} of quarks or leptons. And, this flavor doublet \emph{structure} imposed by the 2-space, necessarily, severely restricts the possible \emph{interactions} of fundamental fermions \cite{52}. For example, we know from the 2-space description that the magnitude of the difference in electric charges between the ``up'' and ``down'' flavors within a flavor doublet is the fundamental unit of electric charge. And, if there are transitions that change an ``up'' flavor to a ``down'' flavor or vice versa, there must exist a weak intermediate boson (with unit electric charge) to accomplish this. Thinking along these lines soon leads to the notion of weak isospin, and $SU(2)_L$ flavor doublets carrying weak isospin. Fourth and finally, because the electric charge of flavor-doublet members is ``explained'' by the 2-space description, compatibility with electromagnetism, i.e., there is only \emph{one} photon associated with flavor-doublet members (see Ref.\ 3, pp.\ 342--343), inevitably leads to the concept of weak hypercharge and $U(1)_Y$. So, while the 2-space certainly does \emph{not} rigidly determine these things, it does seem to severely limit the possibilities.

In this very general way, it seems that the 2-space, and especially whatever underlying physics is responsible for it, ultimately \emph{determines} the current form assumed by the standard model of particle physics based on $SU(3)_\color\times SU(2)_L\times U(1)_Y$. However, until the underlying dynamics responsible for the 2-space is fully understood, the foregoing connections will remain speculative in nature.

\section{Summary and Conclusions}

It is widely believed that the explanation for fundamental-fermion (quark and lepton) flavors, flavor doublets, and family replication is to be found in theories of quantum gravity (e.g., superstrings). And, yet, as demonstrated in this paper and elsewhere [10], an analytic continuation of a Hermitian matrix used to represent the fermion-number operator, leads to a new \emph{internal} description of quarks and leptons---a ``layer'' of new physics located \emph{somewhere} between the unification scale, and the region of applicability of the standard model of particle physics---that also explains such things as family replication. The new description [9--13] involves an (internal) 2-space, and two new associated 2-vector fermion observables $\U$ and $\V$ (linearly-independent flavor-defining ``coordinates'') that are somewhat analogous to the spin coordinate. And, just as the spin coordinate explains certain subtleties of atomic spectra (e.g., the fine structure) these new 2-vector coordinates explain certain subtleties of fundamental-fermion ``spectra.'' In particular, using the two new flavor-defining fermion coordinates $\U$ and $\V$, we have discovered, among other lesser things, that:
\medskip

\noindent 1) The so-called ``accidental'' symmetries of the Lagrangian describing strong and electroweak interactions (of fundamental fermions) are \emph{not} accidental, but instead seem to be imposed on these Lagrangians by an underlying non-Euclidean 2-space---or whatever underlying dynamics is responsible for the 2-space, its quantization, and its associated selection rules. In particular, a central conclusion of this work is this:
\medskip

\noindent\emph{The global} (\emph{flavor-defining}) \emph{charges associated with the ``accidental symmetries'' of the Lagrangian describing strong and electroweak interactions of fundamental fermions, and the global} (\emph{flavor-defining}) \emph{charges associated with the non-Euclidean 2-space, are} (\emph{essentially}) \emph{one and the same charges.}
\medskip

\noindent 2) Family replication is apparently little more than the number of different physically acceptable ways---just \emph{three} ways are predicted---that the ``electric-charge'' vector coordinate $\Q$, for both quark and lepton flavor doublets, can be \emph{resolved} in the 2-space into pairs of linearly independent 2-vector coordinates $\U$ and $\V$ (i.e., $\Q=\U+\V$). However, while this description is reasonable, the underlying reasons for this ``quantization'' remain essentially unknown.
\medskip

\noindent 3) Unlike the situation in the standard model, particles such as the $u$, $c$ and $t$ quarks are characterized by certain associated internal \emph{topological} differences. Similar topological differences between different neutrino flavors may help explain recent observations of (nearly) bi-maximal $\nu_\mu-\nu_\tau$ mixing. And, this in turn, \emph{may be yet another indirect experimental indication of new dynamics} (\emph{related, in part, to topology}) \emph{underlying the new 2-space description of quarks and leptons}.
\medskip

Finally, because the (global) accidental symmetries of the Lagrangian are at least partly determined by the 2-space, and because local symmetries such as $SU(3)_\color$, $SU(2)_L$ and $U(1)_Y$ are also more-or-less \emph{implicit} in the 2-space description, this description tends to imply the standard model of particle physics based on $SU(3)_\color\times SU(2)_L\times U(1)_Y$. However, while progress has been made in making these connections, and in understanding the bewildering number of fermion flavors, two of the most important questions remain unanswered. First, what is the detailed nature of the ``layer'' of new physics, which is supposed to be responsible for the 2-space, its quantization, and its associated selection rules? Second, how does this ``layer'' of new physics arise or \emph{emerge} from physics at the unification scale?

\section{Acknowledgments}

I wish to thank R.\ Zannelli for numerous helpful discussions, and for encouraging me to include footnote \cite{25} on the possible existence (or nonexistence) of exotic fractionally-charged ``quarks.'' I also wish to thank M.\ Sheetz for skillful manuscript preparation, and for submitting the manuscript to the e-Print archive.

\section{References and Footnotes}
\vskip-.2in

\renewcommand{\refname}{}


\begin{thebibliography}{99}
\bibitem[16]{16}  When weak interactions are ``turned off,'' flavor eigenstates and mass 
eigenstates are one and the same.  For the most part, when we speak here of flavor eigenstates, we are referring to the situation where flavor eigenstates
and mass eigenstates are the same.
\bibitem[17]{17}Not counting spin states, we have 3 quark flavor doublets in three colors each, which yields $3\times 2\times 3=18$ states, while 3 lepton flavor doublets in one ``color'' each (i.e., the strong-color ``multiplicity'' is one), yields $3\times 2\times 1=6$ states for a total of 24 matter particles. Combining these 24 matter particles with their 24 (distinguishable) antimatter counterparts, yields a total of 48 fundamental fermions and antifermions. Not counting spin or color states, we have a total of 12 flavor eigenstates and 12 antiflavor eigenstates, for a total of 24 flavor and antiflavor eigenstates of fundamental fermions.
\bibitem[22]{22}Even though $\fv_{\diag}=\fop=\sigma_z$, these two matrices act on entirely \emph{different} spaces, since $\fop$ is associated with the \emph{constant} $\cos\theta=1$, and the \emph{variable} phase-factor $e^{-i\phi}$, whereas $\fv$ is associated with the \emph{variable} $\cos\theta=\cosh v\ge 1$,  and the \emph{constant} phase-factor 
$e^{+
i\phi}=+i$.
\bibitem[34]{34}Given the 2 by 2 matrix $\R$ in Eq.\ (51) in the main text, we see immediately that the effect of $\R$ on the column vectors $\{1/2, -1/2\}$, $\{1,0\}$ and $\{0,-1\}$ is simply $\R\{1/2,-1/2\}=\{1,0\}$ and $\R\{1,0\}=\{0,-1\}$, or $\R^2\{1/2, -1/2\}=\{0,-1\}$.
\bibitem[39]{39}The baryon (lepton) asymmetry of the universe insures that $B$ ($L$) are not absolutely conserved. However, violation of these conservation laws is very strongly suppressed as evidenced by the long life of the proton.
\bibitem[51]{51}Even if the matrix $M$ in Eq.\ (84) in the main text is verified by experiment, we could not be certain that the constraints expressed by  items A), B) and C) in Section 3.5.2, are \emph{topological} constraints related to the 2-space description of quarks and leptons. There are three basic reasons for this assertion.

\qquad First, we do not know why the topology of vector triads [with respect to the internal transformation $\fv$] should be \emph{relevant} to neutrino mixing. Second, we do not have any detailed understanding of the hypothetical \emph{mechanism} that supplies the ``energy barriers'' or ``topological charge conservation laws'' that serve to inhibit topology change in neutrino mixing. Third, as indicated in Ref.\ 13, Appendix A (Sec.\ A.4) in this list, we cannot be absolutely certain that the $\nu_\mu$ neutrino  is associated with the requisite second-family 
M\"obius topology with respect to $\fv$. 
What we do know with certainty is that, regardless of their physical origins, the  constraints of Section 3.5.2, together with the additional assumption that $\sin^2\delta\ne 1$ or $\cos^2\delta\ne 0$, definitely determine $M$ \emph{uniquely}.
\end{thebibliography}
\end{document}